\documentclass[11pt,graphicx,amsmath]{article}
\usepackage{amsmath}
\usepackage{graphicx}
\usepackage{bm}
\usepackage{color}
\usepackage{amssymb}
\usepackage{amsfonts}
\usepackage{comment}
\usepackage{cite}
\usepackage{todonotes}
\usepackage{caption}
\usepackage{subcaption}
\usepackage{appendix}
\usepackage{epstopdf}

\def\be{\begin{equation}}
\def\ee{\end{equation}}
\def\nn{\nonumber}

\def\ba{\begin{eqnarray}}
\def\ea{\end{eqnarray}}
\def\bl#1\el{\begin{align}#1\end{align}}

\def\be{\begin{equation}}
\def\ee{\end{equation}}
\def\ba{\begin{eqnarray}}
\def\ea{\end{eqnarray}}
\def\nn{\nonumber}
\def\bl#1\el{\begin{align}#1\end{align}}

  \title{Field Theory of the Correlation Function of
   	Mass Density Fluctuations for Self-Gravitating Systems}

 \author{\small
            \,  Yang  Zhang\thanks{yzh@ustc.edu.cn} , \,
             Qing Chen  \thanks{cqpb@mail.ustc.edu.cn}  ,\,
                Shu-Guang Wu  \thanks{wusg@mail.ustc.edu.cn}         \\
 \small  Department of  Astronomy,
   CAS Key Laboratory for Researches in Galaxies and Cosmology, \\
 \small  University of Science and Technology of China, Hefei, Anhui, 230026, China \\
 }

 \date{}

\def\be{\begin{equation}}
\def\ee{\end{equation}}
\def\ba{\begin{eqnarray}}
\def\ea{\end{eqnarray}}
\def\nn{\nonumber}
\newcommand{\bee}{\begin{eqnarray}}
\newcommand{\een}{\end{eqnarray}}

\begin{document}

\maketitle

\begin{abstract}\Large

The mass density distribution of Newtonian self-gravitating systems is studied analytically
in field theoretical method. Modeling the system  as a fluid in hydrostatical equilibrium, we apply
Schwinger's functional derivative on the average of the field equation of  mass density, and obtain
the field equation of 2-point correlation function $\xi(r)$ of the mass density fluctuation, which
includes the next order of nonlinearity beyond the Gaussian approximation. The 3-point correlation
occurs hierarchically in the equation, and is  cut off by the Groth-Peebles anzats, making it closed.
We perform renormalization, and write the equation with three nonlinear coefficients. The equation
tells that $\xi $ depends on the point mass $m$ and the Jeans wavelength scale $\lambda_{0}$, which
are different for galaxies and clusters. Applying to large scale structure,	it predicts that the
profile of $\xi_{cc} $ of clusters is similar to $\xi_{gg}$	of galaxies but with a higher amplitude,
and that the correlation length increases with the mean separation between clusters, i.e, a scaling
behavior $r_0\simeq 0.4d$. The solution yields the galaxy correlation $\xi_{gg}(r) \simeq (r_0/r)^{1.7}$
valid only in a range $1<r<10 \,h^{-1}$Mpc.	At larger scales the solution $\xi_{gg} $ deviates below
the power law and goes to zero around $\sim 50 \, h^{-1}$Mpc, just as the observations show. We also
derive the field equation of 3-point correlation function in Gaussian approximation and its analytical
solution, for which the Groth-Peebles ansatz with $Q= 1$ holds.

\end{abstract}

cosmology: large-scale structure of Universe --- cosmology: theory

\large

\section{Introduction}

To understand the  matter distribution in Universe on large scales is
one of the major goals of modern cosmology.
The large scale structure is determined by self gravity of matter.
Since the number of galaxies as the typical objects  is enormous,
one needs statistics to study the distribution.
In this regard,
the $2$-point correlation function $\xi_{gg}(r)$ of galaxies
and $\xi_{cc}(r)$ of clusters
serve as a powerful tool (\cite{Bok(1934),TotsujiKihara(1969),Peebles(1980)}).
It  not only provides the statistical information,
but also contains the underlying dynamics  due to gravity.
Observational surveys have been carried out
for galaxies and for clusters,
such as the Automatic Plate Measuring (APM) galaxy survey
(\cite{Lovedayetal(1996)}),
the Two-degree-Field Galaxy Redshift Survey (2dFGRS) (\cite{Peacocketal(2001)}),
Sloan Digital Sky Survey (SDSS) (\cite{Abazajian et al.(2009)}), etc.
All these surveys suggest that the correlation of galaxies
has a  power law form $\xi_{gg}(r)\propto (r_0/r)^{\gamma}$ with $r_0\sim 5.4 \, h^{-1}$Mpc
and $\gamma\sim 1.7$ in a range $ (0.1\sim 10) \, h^{-1}$Mpc
(\cite{TotsujiKihara(1969),GrothPeebles(1977),Peebles(1980),GrothPeebles(1986),SoneiraPeebles(1978)}).
The correlation of clusters is found to be
of a similar form:  $\xi_{cc}(r)\sim 20 \xi_{gg}(r)$
in a range $ (5\sim 60) \, h^{-1}$Mpc,
with an amplified magnitude (\cite{BahcallSoneira(1983),KlypinKopylov(1983)}).
For quasars $\xi_{qq}(r)\sim 5 \xi_{gg}(r)$ (\cite{Shaver(1988)}).

On theoretical side,
numerical computation is the most used method
and significant progresses have been made in study of the large scale structure.
On the other hand,  analytical studies are also important
in understanding the physical mechanism underlying  the clustering.
Ref. (\cite{Saslaw(1968),Saslaw(1969),Saslaw(1985),Saslaw(2000)})
used macroscopic thermodynamic variables,
such as internal energy, entropy, pressure, etc, for adequate descriptions,
whereby the power-law form of $\xi_{gg}(r)$
was used to calculate modifications to energy and pressure.
Similarly, Ref. (\cite{deVegaetal(1996a),deVegaetal(1998)})  used
the grand partition function of  a self-gravitating gas
to study a possible fractal structure of
the correlation function of galaxies.
However the field equation of $\xi$ was not given in all these studies so far.

In Ref. (\cite{Zhang(2007)}) we studied
mass density distribution of self-gravitating systems in hydrostatical equilibrium
by a field-theoretical method.
The starting point is the field equation of the mass density field $\psi$.
We  expand  the  density field as $\psi=\psi_0+\delta\psi$,
where $\psi_0=\langle \psi  \rangle $ is  the mean density
and  $\delta\psi$ is the fluctuation field.
We employed the technique of the generating functional $Z[J]$
as a path integral over the field $\psi$,
where  $J$ is the external source.
The connected   2-point correlation function  is
$G^{(2)}({\bf r}_1,{\bf r}_2)=\langle \delta\psi( {\bf r}_1)\delta\psi( {\bf r}_2) \rangle
= \delta^2 \ln Z[J] /\delta J ( {\bf r}_1)\delta J( {\bf r}_2)$,
also denoted by $\xi(r)$ with $r=| {\bf r}_1 - {\bf r}_2|$
(\cite{Zhang(2007),ZhangMiao(2009)}).
By taking functional derivative $\delta/\delta J$ of the equation of $\psi_0$,
the field equation of $G^{(2)}$
was derived in Gaussian approximation with nonlinear terms
 of  $\delta\psi$ being neglected.
The analytic solution of correlation function
contains the Jeans wavelength as the unique scale of self-gravitating systems,
and, as a prominent property,
the amplitude of the correlation is proportional to the mass of particle.
This  feature explains naturally
the observational fact that  clusters have a correlation amplitude higher than galaxies,
and, similarly, richer clusters has a correlation amplitude higher than poor ones.
When applying to large scale structure,
the solution agreed qualitatively with
the observed correlation of galaxies and of clusters,
however,  at small scales $r<3 \, h^{-1}$Mpc,
the correlation is too low to account for
what is observed.

To improve the Gaussian approximation,
Ref. (\cite{ZhangMiao(2009)}) considered nonlinear terms of density fluctuation
to order of $(\delta\psi)^2$
and gave the nonlinear equation of $G^{(2)}$.
Due to hierarchy,
the equation contains  the 3-point correlation $G^{(3)}$,
which can be expressed as the products of $G^{(2)}$
by the Kirkwood-Groth-Peebles ansatz (\cite{Kirkwood(1932),GrothPeebles(1977)})
leading to the closed equation of $G^{(2)}$.
After necessary renormalization to absorb
the quantities like  $G^{(2)}(0)$ etc,
the nonlinear  equation  was obtained.
The  correlation is  enhanced at small scales $r=(0.3 \sim 3) \, h^{-1}$Mpc,
substantially improving the Gaussian result.
But the treatment is not complete, as a  nonlinear  term
was not properly included.

This paper extends the previous preliminary work (\cite{Zhang(2007),ZhangMiao(2009)})
by a complete treatment of all  terms  $(\delta\psi)^2$,
and presents the detailed derivation of the field equation of $G^{(2)}$ and
renormalization procedure.

Besides, this paper also presents
the field equation of 3-point correlation $G^{(3)}$
in Gaussian approximation.
These  will complete the work of Ref. (\cite{ZhangChen(2015)}),
which listed only  the brief results  on $G^{(2)}$ without details.
With one  set of fixed values of nonlinear coefficients,
the solution $G^{(2)}$ of resulting field equation
will confront the observational data of both galaxies and clusters.

In section \ref{sect:equSGS}, we derive the field equation of density field $\psi $ by hydrostatics,
and write down the generating functional $Z[J]$.

Section \ref{sect:equ2pt} is an outline the  derivation of  the nonlinear field equation of $\xi(r)$,
using the functional derivative technique.

Section \ref{Sect:GPofFE} gives the main predictions by the equation on the properties of clustering.

In section \ref{Sect:ApptoGa}, we present the solution $\xi(r)$
to confront with observations of galaxies,
and compare with numerical simulations as well.
We also give  the projected correlation function.

In section \ref{Sect:ApptoCl}, we apply the same solution to the system of clusters with greater mass $m$.

Section \ref{Sect:3PTCF} gives the 3-point correlation function $G^{(3)}$ at Gaussian approximation.

Section \ref{Sect:ConDisc} contains conclusions and discussions.

Appendix \ref{App:GPF} gives the functional $Z[J]$
of the many-body self-gravitating system in terms of path integral
over the gravitational field.

Appendix \ref{App:FEandRenorm} presents the  details of the derivation
of the field equation of $\xi(r)$,
including the use of  Kirkwood-Groth-Peebles ansatz
and the renormalization procedure.

Appendix \ref{App:3PTCF} gives the derivation of
the field equation of $G^{(3)}$ at  Gaussian approximation.

We use a unit in which the speed of light $c=1$ and the Boltzmann constant $k_B=1$.

\section{Field Equation of Mass Density of Self-Gravitating System }
\label{sect:equSGS}

Galaxies, or clusters, distributed in Universe
can be approximately described
as a fluid at rest in the gravitational field,
i.e, as self-gravitating hydrostatics.
This modeling is  an approximation
since the cosmic expansion is not considered.
The system of galaxies in the expanding Universe
is in an asymptotically relaxed state,
i.e, a quasi thermal equilibrium (\cite{Saslaw(2000)}).
In this paper,
under the approximation of hydrostatical equilibrium,
we study the system of galaxies within a small redshift range.
Let us examine how far this approximation is from the actual situation.
The  time scale of the cosmic expansion is $t_e \equiv 1/H_0 = \sqrt{3/8\pi G \rho_0}$,
and the dynamical time for galaxies moving in
the background is $t_d\sim \sqrt{3/16 \pi G \rho_0}$  (\cite{BinneyTremaine(1987)}),
and the  two time scales are roughly of the same order of magnitude,
so the hydrostatical  equilibrium is not a bad approximation,
as will be demonstrated further in Section \ref{Sect:ApptoGa}.

In general,
a fluid is described by the continuity equation, the Euler equation,
and the Poisson equation:
\be
\frac{\partial \rho}{\partial t} +\nabla \cdot (\rho {\bf v})=0,
\ee
\be
\frac{\partial \bf v}{\partial t}+({\bf v}\cdot \nabla){\bf v}
=-\frac{1}{\rho}\nabla p + \nabla \Phi,
\ee
\be \label{poisson}
\nabla^2 \Phi =-4\pi G \rho.
\ee
For the hydrostatical case, $\dot\rho=0$ and ${\bf v}=0$,
the Euler equation takes the form (\cite{LandauLifshitz(1987)})
\be \label{Eulereq2}
\frac{1}{\rho}\nabla \rho = \frac{1}{c_s^2}\nabla \Phi,
\ee
with $c_s^2 \equiv \partial p/\partial \rho$  being a constant sound speed,
which describes the mechanical equilibrium of the fluid.
Taking gradient on both sides of this equation
and using Eq. (\ref{poisson}) and (\ref{Eulereq2})
leads to
\be  \label{rederivation}
\nabla^{2}\rho-\frac{1}{\rho}(\nabla\rho)^{2}+\frac{4\pi G}{c_s^2}\rho^2=0.
\ee
We call Eq. (\ref{rederivation}) the field equation of  mass density
for the self-gravitating fluid system.
For convenience, we introduce a dimensionless density field
$\psi ({\bf r}) \equiv \rho({\bf r})/\rho_{0}$,
where  $\rho_0=mn_0$ is the mean mass density of the system.
Then Eq. (\ref{rederivation}) takes the form
\be \label{masseq}
\nabla^{2}\psi-\frac{1}{\psi}(\nabla\psi)^{2}+k_{J}^{2}\psi^{2}=0,
\ee
with $k_{J}\equiv \sqrt{4\pi G\rho_{0}}/c_{s}$ being the Jeans wavenumber.
This is highly nonlinear in  $\psi$ as it contains $1/\psi$.
Eq. (\ref{masseq}) also follows from $\delta\mathcal{H}(\psi)/\delta \psi=0$
with the effective Hamiltonian density
\be \label{Hpsi}
\mathcal{H}(\psi)=\frac{1}{2}(\frac{\nabla\psi}{\psi})^{2}-k_{J}^{2}\psi.
\ee
To employ
Schwinger's technique of functional derivatives (\cite{Schwinger(1951a), Schwinger(1951b)}),
an external source $J(\bf r)$ is introduced to coupled with the  field $\psi$:
\be  \label{eff_L}
\mathcal{H}(\psi,J)=\frac{1}{2}(\frac{\nabla\psi}{\psi})^{2}
-k_{J}^{2}\psi-J\psi,
\ee
and the mass density field equation in the presence of $J$ is
\[
\frac{1}{\psi^2}  \nabla^{2}\psi-
\frac{1}{\psi^3}  (\nabla\psi)^{2}+k_{J}^{2} +J =0.
\]
When $\psi\ne 0$ and $\psi\ne \infty$, one has
\be  \label{field_eq}
\nabla^{2}\psi-\frac{1}{\psi}(\nabla\psi)^{2}+k_{J}^{2}\psi^{2}+J\psi^{2}=0.
\ee
This is the starting equation which we shall use to derive the field equation of
2-point correlation function $G^{(2)}(r)$.
When $\psi\ne 0$ and $\psi\ne \infty$,
the generating functional for
the correlation functions of $\psi$ is defined as
\be \label{ZJ}
Z[J] = \int  D\psi
e^{-\alpha \int d^{3}\textbf{r}\mathcal{H}(\psi,J)},
\ee
where $\alpha  \equiv c_s^2/4\pi G m$ with $c_s $ being the sound speed
and $m$ being the mass of a single particle.

By setting $\psi( {\bf r})\equiv e^{\phi( {\bf r})}$,
where $\phi \equiv \Phi/ c_s^2$ is the gravitational potential,
Eq. (\ref{masseq}) and (\ref{Hpsi}) can also be transformed into
the well-known Lane-Emden equation
(\cite{Emden(1907),Ebert(1955,Bonnor(1956),Antonov(1962),Lynden-BellWood(1968)}):
\be \label{Lane-Emden}
\nabla^2\phi + k^2_J e^{\phi}=0,
\ee
and
the associated  Hamiltonian
\be \label{L}
\mathcal{H}(\phi)= \frac{1}{2 }(\nabla \phi)^2-k_J^2e^{\phi} .
\ee
In fact,
Eq. (\ref{L}) can also derived from the following approach.
The Universe filled with galaxies and clusters
can be modeled as a self gravitating gas
assumed to be in thermal quasi-equilibrium (\cite{Saslaw(1985),Saslaw(2000)}).
Note that the Universe is expanding with a time scale  $\sim 1/H_0 =(3/8\pi G \rho_0)^{1/2}$,
and the time scale of propagation of fluctuations
$\sim \lambda_J/c_s\sim 1/(4\pi G \rho_0)^{1/2}$,
both being of the same order of magnitude.
The thermal equilibrium is an approximation.
For such a system of $N$ particles of mass $m$,
the Hamiltonian is
\be \label{Hamiltonian}
H=\sum_{i=1}^N \frac{p_i^2}{2m}-\sum_{i<j}^N \frac{Gm^2}{r_{ij}}
\ee
with $r_{ij}=|{\bf r}_i- {\bf r}_j|$,
and the grand partition function
is
\be \label{Z1}
Z=\sum_{N=0}^\infty \frac{z^N}{N!}
\int \prod_{i=1}^N  \frac{d^3p_i\, d^3r_i}{(2\pi)^3}
e^{-H/T},
\ee
where $z$ is the fugacity.
As shown in Appendix \ref{App:GPF},
this can be written as a path integral over the field $\phi$:
\be\label{Zphi}
Z=\int D \phi \,
e^{-\alpha\int d^3 r  \mathcal{H}(\phi)  },
\ee
where $\mathcal{H}(\phi)$ is given in Eq. (\ref{L}) with $c^2_s=T/m$.
In this paper, we shall use Eq. (\ref{masseq}) and (\ref{Hpsi}),
which suit better for studying the mass distribution.

\section{Field Equation of the 2-point Correlation Function of density fluctuations}
\label{sect:equ2pt}

\begin{figure}
	\includegraphics[width=14.0cm, angle=0]{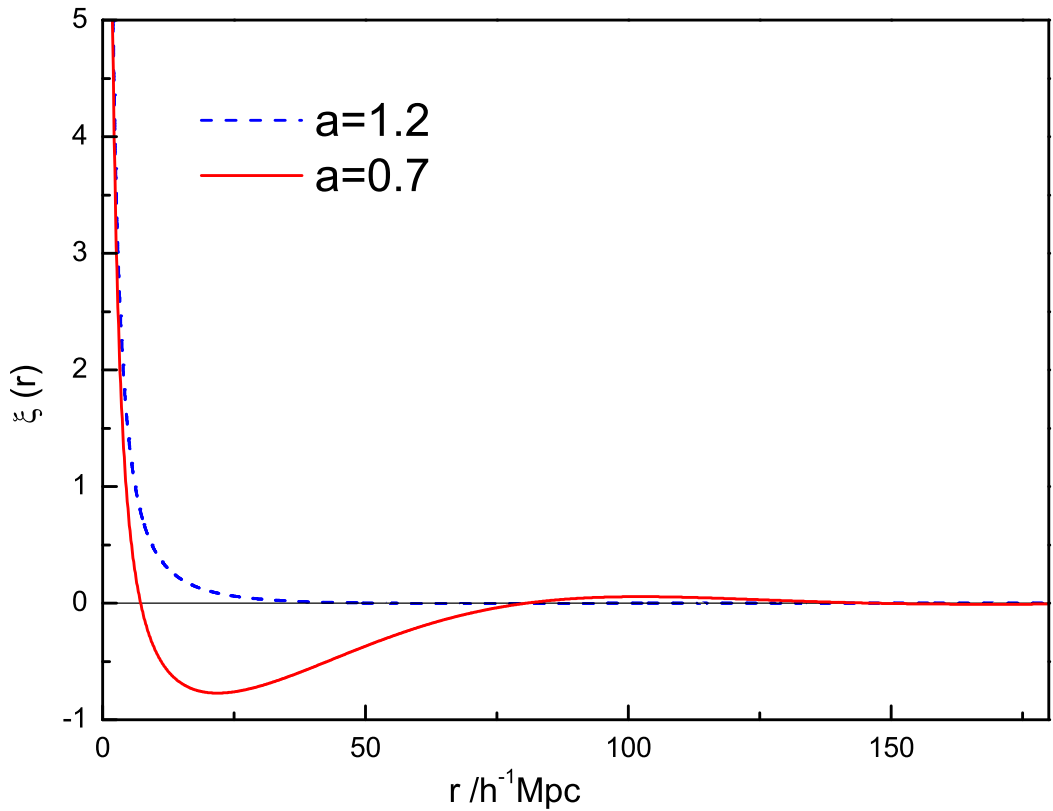}
	\caption{
		A large viscosity coefficient $a$
		will cause strong damping to the oscillations of $\xi(r)$
		at large distances.
		In this graph,  $b$, $c$, and $k_0$
		are fixed for demonstration.
	}
	\label{a}
\end{figure}

In this section we outline the field equation of 2-point correlation function
of density fluctuations,
and the comprehensive details are attached in Appendix \ref{App:FEandRenorm}.
Consider the fluctuation field
$\delta\psi(\bf{r}) \equiv  \psi(\textbf{r})-\langle\psi(\textbf{r})\rangle$,
with the mean
\begin{align}
\langle\psi(\textbf{r})\rangle =
\frac{1}{Z}\int D\psi \,\psi e^{-\alpha \int d^{3}\textbf{r}\mathcal{H}(\psi)}
=\frac{\delta \log Z[J] }{\alpha \delta J({\bf r })} \mid_{J=0},
\end{align}
where one sets $J=0$ after taking functional derivative.
In a general system,
the mean density $\langle\psi(\textbf{r})\rangle $
can vary in space,
but for the homogeneous and isotropic Universe it is a constant $\langle\psi(\textbf{r})\rangle=\psi_{0}$.
The 2-point correlation function of $\delta \psi$,
i.e, the {\it connected} 2-point Green function,
is given by the functional derivative of $\log Z[J]$
with respect to $J$ (\cite{Binneyetal(1992)}) :
\begin{align}\label{2ptGreen}
G^{(2)}({\bf r}_{1},{\bf r}_{2})
&\equiv \langle\delta\psi({\bf r}_{1})\delta\psi({\bf r}_{2}) \rangle\nonumber\\
&=\alpha^{-2}
\frac{\delta^2}{\delta J({\bf r}_1)\delta J({\bf r}_2 )}
\log Z[J]|_{J=0}\nonumber\\
&=\alpha ^{-1}
\frac{ \delta \langle\psi({\bf r}_{2})\rangle_J }{\delta J({\bf r}_{1})} \, |_{J=0},
\end{align}
where
$\langle\psi(\textbf{r})\rangle_J
\equiv \frac{\delta}{\alpha \delta J({\bf r })}\log Z[J]$
before setting $J=0$.
One can take
$G^{(2)}({\bf r}_{1},{\bf r}_{2}) = G^{(2)}(r_{12}) $
for a homogeneous and isotropic universe.
Analogously, the n-point correlation function of $\delta \psi$ is
\begin{align}
G^{(n)}({\bf r}_{1},..., {\bf r}_{n})
&\equiv \langle\delta\psi({\bf r}_{1})...\delta\psi({\bf r}_{n}) \rangle\nonumber\\
&=\alpha^{-n}
\frac{\delta^n  \log Z[J]}{\delta J({\bf r}_1)...\delta J({\bf r}_n )}
|_{J=0}\nonumber\\
& =\alpha^{-(n-1)}
\frac{\delta^{n-1} \langle\psi({\bf r}_n)\rangle_J}{\delta J({\bf r}_1)
	...\delta J({\bf r}_{n-1} )}   |_{J=0}
\end{align}
for $n\ge 3$.
To derive the field equation of $G^{(2)}(r)$, as a routine (\cite{Goldenfeld(1992)}),
one takes functional derivative
of the ensemble average of Eq. (\ref{field_eq})
with respect to $J({\bf r}')$,
\be  \label{Eqintermediate}
\frac{\delta}{\delta J({\bf r}')}
\bigg(\langle\nabla^{2}\psi({\bf r})\rangle_J
-\langle \frac{1}{\psi({\bf r})}(\nabla\psi({\bf r}))^{2} \rangle_J
+k_{J}^{2} \langle \psi({\bf r})^{2}\rangle_J
+J({\bf r})\langle \psi({\bf r})^{2}\rangle_J\bigg)=0,
\ee
and then sets $J=0$.
To systematically deal with the nonlinearity of $1/\psi$,
we expand $\psi$ in terms of the fluctuation $\delta\psi$,
and keep up to the second order  $(\delta \psi)^2$.
Then Eq. (\ref{Eqintermediate}) leads the following equation of $G^{(2)}$:
\ba  \label{2pt3pt}
&&\nabla^{2}G^{(2)}(\textbf{r})
+k_{0}^{2}\psi_{0}G^{(2)}(\textbf{r})
+\frac{1}{2\psi_{0}^{2}}\nabla^{2}G^{(2)}(0)G^{(2)}(\textbf{r})
-(\frac{1}{2\psi_{0}}\nabla^{2}+k_{J}^{2})G^{(3)}(0,{\bf r},{\bf r})\nonumber\\
&&+\frac{2}{\psi_{0} ^{2}}\nabla G^{(2)}(0)\cdot\nabla G^{(2)}(\textbf{r})
=-\frac{1}{\alpha }\big( \psi_{0}^{2}-G^{(2)}(0) \,\big) \delta^{(3)}(\textbf{r}),
\ea
where the characteristic wavenumber  $k_0\equiv \sqrt{2}k_J$.
(See Appendix \ref{App:FEandRenorm} for detailed calculations.)
This equation is of the same form as
Eq. (4) in our previous paper (\cite{ZhangMiao(2009)}),
except that we now keep $-k_{J}^{2}G^{(3)}$ on r.h.s,
and $\frac{1}{\alpha }G^{(2)}(0)\delta^{(3)}(\textbf{r})$ in the source on l.h.s.
These new terms come from
an improved treatment which include high order contributions properly.
Note that $G^{(3)}$ occurs in Eq. (\ref{2pt3pt}).
There are various ways to cut off this hierarchy.
And one of them is to use the Kirkwood-Groth-Peebles ansatz
 (\cite{Kirkwood(1932),GrothPeebles(1977)})
\be \label{Ansatz}
G^{(3)}(\textbf{r}_1,\textbf{r}_2,\textbf{r}_3)
=Q\left(G^{(2)}(r_{12} ) G^{(2)}(r_{23} )
+G^{(2)}(r_{23} ) G^{(2)}(r_{31} )
+G^{(2)}(r_{31} ) G^{(2)}(r_{12} )\right),
\ee
where $Q$ is a dimensionless parameter.
The observational data of galaxy surveys
have indicated that $Q\simeq 1\pm 0.2$,
which is also supported by numerical simulations (\cite{Peebles(1993)}).
Here we adopt this ansatz.
Substituting Eq. (\ref{Ansatz}) into Eq. (\ref{2pt3pt}),
after a necessary renormalization to absorb
the quantities like  $G^{(2)}(0)$,  $\nabla G^{(2)}(0)$, and $\nabla^2 G^{(2)}(0)$,
we obtain the field equation of 2-point correlation function
\begin{align} \label{eqfinal}
&(1-b\xi)\nabla^{2}\xi
+k_{0}^{2}(1- c\xi)\xi
+({\bf a}-b\nabla \xi) \cdot\nabla \xi
=-\frac{1}{\alpha }  \delta^{(3)}(\textbf{r}),
\end{align}
where
$ \xi(r) \equiv  G^{(2)}(\textbf{r})$,
and  $\bf a$, $b$, and $c$ are three constant parameters.
The special case of ${\bf a}=b=c=0$  is the Gaussian approximation,
and Eq. (\ref{eqfinal}) reduces to
the Helmholtz equation (\ref{Helmoltz}).
Thus, the terms of $\bf a$, $b$, and $c$
represent the nonlinear contributions beyond
the Gaussian approximation.
Eq. (\ref{eqfinal}) in the radial direction is
\be \label{final}
(1-b\xi)\xi''+
( (1-b\xi)\frac{2}{x}+a )\xi' +\xi  -b \xi'\, ^{2}   -c\xi^{2}=
-\frac{1}{\alpha }   \frac{\delta(x)k_0}{x^{2}},
\ee
where
$\xi'\equiv \frac{d}{dx}\xi$, $x \equiv k_{0}r$ and $k_0=(8\pi G m n)^{1/2}/c_s$.
The effects of nonlinear terms of $b$ and $c$
can enhance the amplitude of $\xi$ at small scales
and increase the correlation length.
The term of $a$ plays the role of effective viscosity,
and a greater $a$ leads
a strong damping to the oscillations of $\xi$ at large scales,
as shown in Fig. \ref{a} .
The value of $a$ should be large enough
to ensure  $1+\xi(r)  \ge 0 $ for the whole range  $0<r<\infty$.
The solution $\xi(r)$ will confront
the observational data of galaxies and clusters
in sections \ref{Sect:ApptoGa} and \ref{Sect:ApptoCl}.

\section{General Predictions of  Field Equation}
\label{Sect:GPofFE}

Inspection of Eq. (\ref{eqfinal}) already
reveals its predictions of the important properties of correlation.

1) Eq. (\ref{eqfinal}) can apply to the system of galaxies,
as well as to the system of clusters,
and the only  difference is their respective $m$ and $k_0$ contained in the equation.
Thus, the  solutions of Eq. (\ref{eqfinal})
for galaxies  have a profile similar to that for  clusters.
This explains  the observational fact
that the correlation functions of galaxies and of clusters
have the same power-law form,
$\xi_{gg}, \xi_{cc}\propto r^{-1.8}$,
but different amplitudes and ranges (\cite{BahcallSoneira(1983),KlypinKopylov(1983)}).

2) The $\delta^{(3)}(\bf r)$ source of  Eq. (\ref{eqfinal}) is proportional to
$ 1/\alpha =4\pi G m /c_s^2$,
which determines the overall amplitude of a solution $\xi$.
The sound speed  $c_s$ can be regarded as the the peculiar velocity,
which is the same order of magnitude
for galaxies and clusters, around several hundreds km/s
(\cite{Hawkinsetal(2003),Mastersetal(2006)}).
Therefore, $1/\alpha$ is essentially determined by $m$,
and a greater $m$ will yield a higher amplitude
\be \label{xim}
\xi(r)  \propto m.
\ee
This general prediction
naturally explains a whole chain of   observed facts:
luminous galaxies are more massive and have a higher correlation amplitude
than ordinary galaxies (\cite{Zehavietal(2005)}),
clusters are much more massive and have a much higher correlation than galaxies,
and rich clusters have a higher correlation
than poor clusters
(\cite{BahcallSoneira(1983),Einastoetal(2002),Bahcalletal(2003)}).
This phenomenon has been a puzzle for long (\cite{Bahcall(1996)})
and was interpreted as being caused by the statistics
of rare peak events (\cite{Kaiser(1984)}).

3) The power spectrum,
as the Fourier transform of $\xi(r)$,
is proportional to the inverse of the spatial number density:
\be \label{pkn}
P(k) \propto 1/n_0,
\ee
(See Eq. (\ref{Pkgaussian})).
The observed $P(k)$ of clusters is much higher than that of galaxies,
which is explained by Eq. (\ref{pkn})
as $n_0$ of clusters is much lower than that of galaxies (\cite{Bahcall(1996)}).
Since a greater $m$ implies a lower $n_0$ for a given  mean mass density $\rho_0=m n_0$,
the properties (\ref{xim}) and (\ref{pkn})
reflect the same physical law of clustering from different perspectives.

4) The characteristic length $\lambda_0=2\pi/k_0
=(\frac{\pi}{2})^{1/2}\frac{c_s}{\sqrt{G\rho_0}}  \propto \frac{c_s}{\sqrt{\rho_0}}$
appears in Eq. (\ref{eqfinal}) as the only scale,
and  underlies the scale-related features of the solution $\xi(r)$.
Observations reveal that clusters have a longer ``correlation length" than galaxies.
This can  be explained by the following.
At a fixed $\lambda_0$,   $\xi_{cc}(r)$ has a higher amplitude
and drops to its first zero at a larger distance,
leading to an apparently  longer  ``correlation length" than  $\xi_{gg}(r)$.
Another possibility may also contribute to this effect:
if $\rho_0$ of the region covered by cluster surveys
is lower than that of  galaxy surveys,
$\lambda_0 $ for cluster  will be longer accordingly.
As will be seen in the next Section 6 and 7,
to use the  solution
to match the data of both galaxies and clusters,
one has to take  a longer $\lambda_0$ for clusters than for galaxies
(\cite{Collinsetal(2000),BahcallSoneira(1983)}).

\section{Applying to Galaxies}
\label{Sect:ApptoGa}

Now we give the solution $\xi_{gg}(r)$ of Eq. (\ref{final})
for a fixed set of parameters $(a,b,c)$,
and confront with the observed correlation from major galaxy surveys.
In choosing the boundary condition at certain point $r_0 \sim 0.1\, h^{-1}$Mpc,
we  $\xi_{gg}(r_0)>0$ and $\xi'_{gg}(r_0)< 0$,
corresponding to $\cos(k_0 r)/r$  in Eq. (\ref{GGaussian})
for the Gaussian case (\cite{Zhang(2007)}).
This choice is similar to the choice of
the adiabatic  mode in the initial condition
of CMB anisotropies (\cite{HuSugiyama(1995)}).
We shall  also convert $\xi_{gg}(r)$ into
its associated  projected correlation function $w_p(r_p)$, simultaneously.

\begin{figure}
	\includegraphics[width=14.0cm, angle=0]{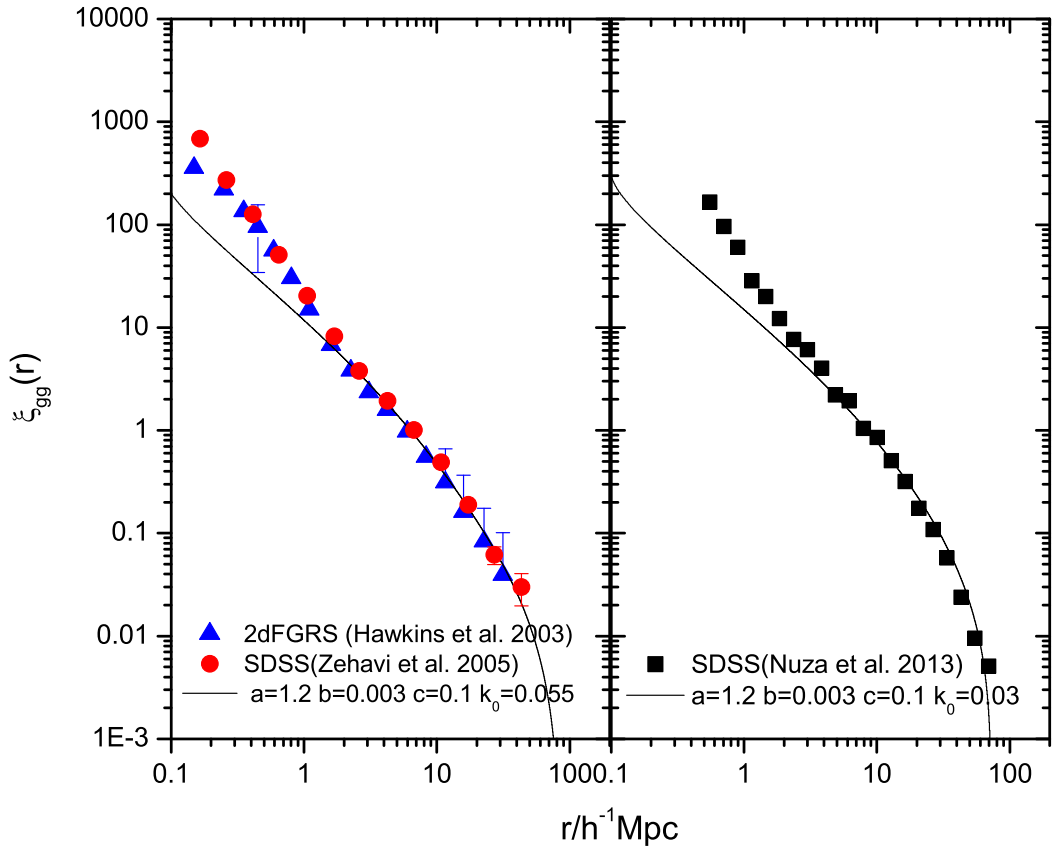}
	\caption{
		Left:
		The solution $\xi_{gg}(r)$ with $k_0=0.055 \,h$Mpc$^{-1}$
		is compared with the data of galaxies by
		2dFGRS having a median redshift $z\sim 0.11$ (\cite{Hawkinsetal(2003)}),
		SDSS having a median $z\sim 0.1$ (\cite{Zehavietal(2005)});
		Right:  The solution $\xi_{gg}(r)$ with $k_0=0.03 \,h$Mpc$^{-1}$
		is compared with  SDSS R9 having a median $z \sim 0.53$ (\cite{Nuzaetal(2013)}).
	}
	\label{correlation}
\end{figure}

First we consider the correlation function $\xi_{gg}(r)$.
For demonstration,
we take the parameters $(a,b,c)=(1.2,0.003, 0.1)$,
though  other values of $(a,b,c)$ can be also chosen.
Fig. \ref{correlation} (Left) shows
the solution $\xi_{gg}(r)$  with  $k_0=0.055 \,h$Mpc$^{-1}$
and the observed  $\xi_{gg}(r)$  by the galaxy surveys
of  2dFGRS (\cite{Hawkinsetal(2003)})
and SDSS (\cite{Zehavietal(2005)}) with a median $z\sim 0.1$.
And Fig. \ref{correlation} (Right) shows the data of SDSS R9 with a median redshift $z\sim 0.53$,
which we have converted from the data of the projected correlation function in Ref. (\cite{Nuzaetal(2013)}).
Since the samples in Right and Left have different $z$
corresponding to different evolution stages,
we have accordingly chosen
two different values $k_0$ and boundary conditions
$\xi_{gg}(r_b)$ and $\xi'_{gg}(r_b)$  to compare with the data.
It is seen that
the theoretical $\xi_{gg}(r)$
matches the observational data.
The power law  $\xi_{gg} \propto r^{-1.7}$ is valid only
in   $r=(0.1 \sim 10) \, h^{-1}$Mpc,
and deviates from both data and  solution   on large scales.
Moreover,  the solution predicts that
$\xi_{gg}(r)$ decreases to zero and becomes negative around $\sim 70 \, h^{-1}$Mpc,
where data are not available currently.
On small scales $r \leq 1 \, h^{-1}$Mpc,
the solution improves the Gaussian approximation (\cite{Zhang(2007)}),
but  is still lower than the data.
This insufficient clustering at $r \leq 1 \, h^{-1}$Mpc
is possibly due to neglect of high-order nonlinear terms   $(\delta \psi)^3$
in perturbations.
Note that Eq. (\ref{eqfinal}) has been derived assuming  $\delta\psi <1$,
and  to extrapolate the solution   $\xi_{gg} $ down  to  smaller  scales
is only an approximation.

\begin{figure}
	\includegraphics[width=14.0cm, angle=0]{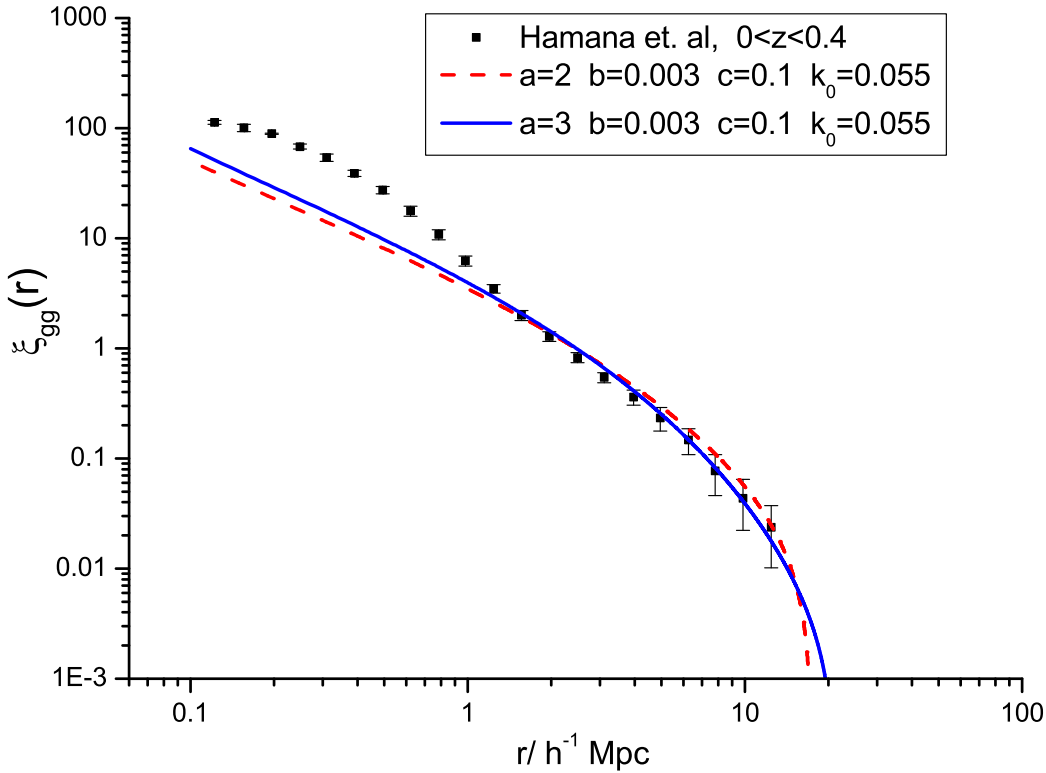}
	\caption{
		The solution $\xi_{gg}(r)$
		is compared with that of the simulations of Hanama et al (\cite{Hamanaetal(2001)}).
		Here  $a= 2,3$ is taken respectively, greater than that used in Fig. \ref{correlation}
		for the survey data.}
	\label{Simulation}
\end{figure}

Here we reexamine the  assumption of hydrostatical equilibrium
in the quasi-linear regime in the expanding Universe.
As is known,  in the quasi-linear regime,
the density fluctuation $\delta\psi  \propto a(t)^{0.3} $ approximately,
where   $a(t)$ is the scale factor in the present stage of accelerating expansion.
So the   time-evolving correlation function
$\xi_{gg}(r,t)=
\langle \delta\psi  \delta\psi  \rangle \propto a^{0.6}(t)=1/(1+z)^{0.6}$.
We have used the static, calculated $\xi_{gg}(r)$
to compare with  the observed correlation function $\xi_{gg}(r,t)$
in an expanding background.
Let us  estimate the errors in doing this.
Take the static $\xi_{gg}(r)$ to correspond to the observed $\xi_{gg}(r,z=0)$.
Within the quasi-linear regime,
the ratio $\xi_{gg}(r)/\xi_{gg}(r,t)  \simeq  (1+z)^{0.6}\simeq 1+0.6 z$
for $z\ll 1$, and the  error is of order $0.6z $.
For the  sample of $\sim 200,000$ galaxies of SDSS (\cite{Zehavietal(2005)}),
the redshift range is $z=(0.02\sim 0.167)$ with a median $z\sim 0.1$.
Take its maximum $z = 0.167$,
and the ratio $\xi_{gg}(r)/\xi_{gg}(r,t) \simeq  (1+0.167)^{0.6} \sim 1.097$,
giving an  error  $0.6z \simeq  0.1 $.
Thus, using the static $\xi_{gg}(r)$ to describe
the time-evolving $\xi_{gg}(r,t)$ of SDDS
has a small  error for $z\ll 1$.
This analysis has also been supported
by studies of numerical simulations.
Ref. (\cite{Hamanaetal(2001)}) has simulated the time-evolving correlation
for $\Lambda$CDM model and demonstrates that
$\xi_{gg}(r,z)$ has changed by a small amount during  $z=0.4 \sim 0$.
The profiles of $\xi_{gg}(r,z=0)$ and  $\xi_{gg}(r,z=0.4)$ are very similar
on a whole range $r= (0.1- 60) \,h^{-1}$Mpc (also similar to our theoretical profile $\xi_{gg}(r)$),
and the ratio
$\frac{ \xi_{gg}(r,z=0)}{\xi_{gg}(r,z=0.4)} \sim  1.3 $ for $r=(5-40) \, h^{-1}$Mpc.
Similar results are also found in other numerical studies
 (\cite{Yoshikawaetal(2001),Taruyaetal(2001)}).
This result agrees
with the estimate based on our analysis in the last paragraph.
Hence, the hydrostatic assumption,
as an approximation, can be applied to the system of galaxies  with  $z\ll 1$,
causing  a small error only.

In actually,
the observed data of  $\xi_{gg}(r)$ are inevitably contaminated by redshift distortions
to various degrees.
Since our analytical solution $\xi_{gg}(r)$  is given in real space,
thus it is more realistic  to compare
our result  directly with those of numerical simulations
in real space that are free of distortions.
Fig. \ref{Simulation} shows that on scales  $r>1 \,h^{-1}$Mpc
our solution  $\xi_{gg}(r)$ agrees very well
with the simulated one given by Hamana et al (\cite{Hamanaetal(2001)}).
Here  the viscosity parameter $a= 2$ or $3$ has been  taken,
greater than $a=1.2$  used in Fig. \ref{correlation}.
Similarly, the insufficiency of amplitude  on small scales $r<1 \, h^{-1}$Mpc
should be improved by including higher order nonlinear terms.

\begin{figure}
	\includegraphics[width=14.0cm, angle=0]{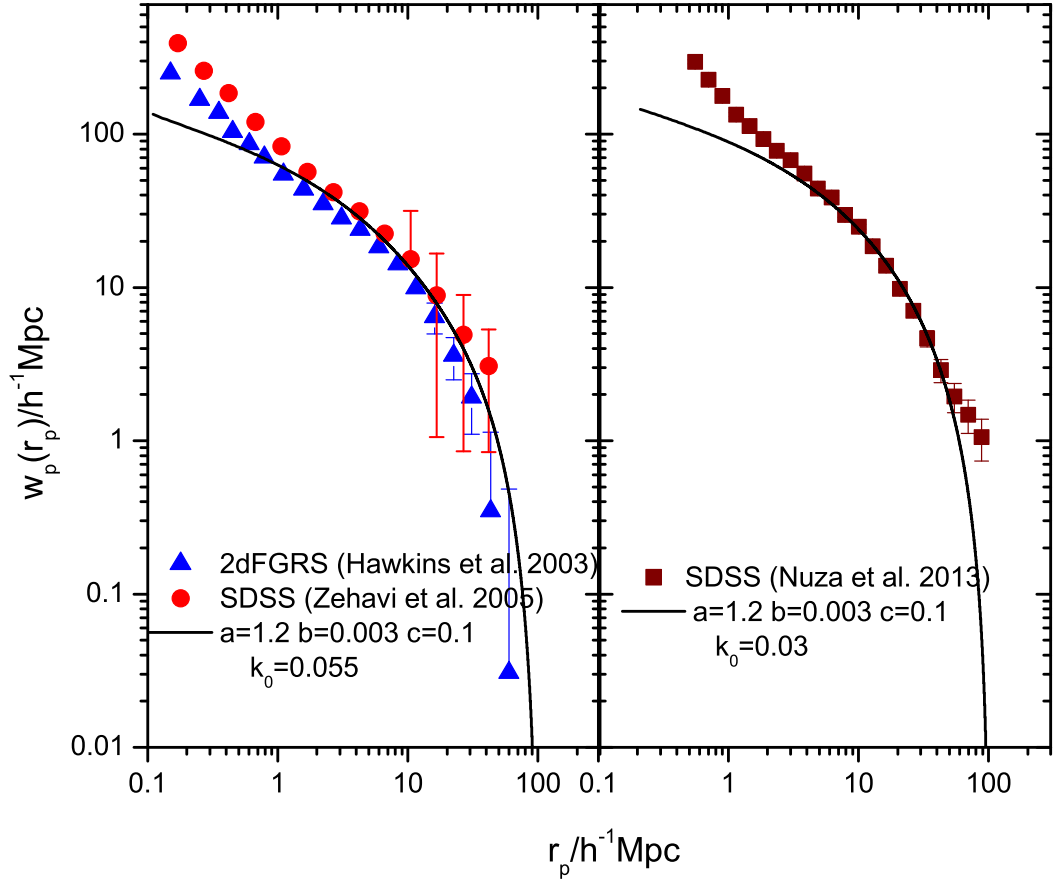}
	\caption{
		The projected correlation function $W_{p}(r_p)$
		converted from  $\xi_{gg}(r)$ confronts
		the data.
		Left: 2dFGRS (\cite{Hawkinsetal(2003)}), SDSS (\cite{Zehavietal(2005)})  with a median $z\sim 0.1$,
		Right: SDSS R9 (\cite{Nuzaetal(2013)}) with a median $z\sim 0.5$.
	}
	\label{Projection}
\end{figure}

Next, we  consider the projected correlation function.
For  sky surveys of galaxies and clusters,
the measurement of distances is through their cosmic red-shift $z$.
Galaxies or clusters have  peculiar velocities,
causing the red-shift distortion to the measured distance.
To eliminate this distorting effect,
one integrates over the distance parallel to the line of sight.
This leads to the projected correlation function (\cite{Peebles(1980)})
\begin{align} \label{real_proj}
W_{p}(r_{p})&=2\int\limits_{0}^{\infty}\xi(\sqrt{r_{p}^{2}+y^{2}})dy
=2\int\limits_{r_{p}}^{\infty}\xi(r)\frac{rdr}{\sqrt{r^{2}-r_{p}^{2}}} \, ,
\end{align}
where $r_{p}$ is the separation of two points
vertical to the line of sight,  not distorted by the peculiar velocities.
In Fig. \ref{Projection} (Left)
the theoretical $W_p(r_p)$ is
converted from the solution $\xi_{gg}(r)$ in Fig. \ref{correlation} (Left),
and compares the observational data of projected correlation function
of 2dFGRS (\cite{Hawkinsetal(2003)}) and SDSS (\cite{Zehavietal(2005)}).
In Fig. \ref{Projection} (Right)
the theoretical $W_p(r_p)$ with $k_0=0.03 \, h$Mpc$^{-1}$
compares  the data of SDSS R9 (\cite{Nuzaetal(2013)}) with a median redshift $z\sim 0.53$.
Overall,
the theoretical $W_{p}(r_{p})$ traces the observational data well
in the range $r_p=(1 \sim 40) \,h^{-1}$Mpc,
but,  is lower than the data
on small scales $r_p\leq 1 \,h^{-1}$Mpc,
the same insufficiency  mentioned before.

\section{Applying to Clusters}
\label{Sect:ApptoCl}

Clusters are believed to trace the cosmic mass distribution
on even larger scales,
and the observational data cover spatial scales that exceed those of galaxies.
Now we  apply
the solution with the same two sets of  $(a,b,c)$ as in Section \ref{Sect:ApptoGa}
to the system of clusters.
Clusters have a greater mass $m$ than that of galaxies,
leading to a higher overall amplitude of $\xi_{cc}(r)$.
In addition,  to match the observational data of clusters,
a small value $k_0=0.03 \,h$Mpc$^{-1}$ is required,
which is lower  than that for galaxies.
\begin{figure}
	\includegraphics[width=14.0cm, angle=0]{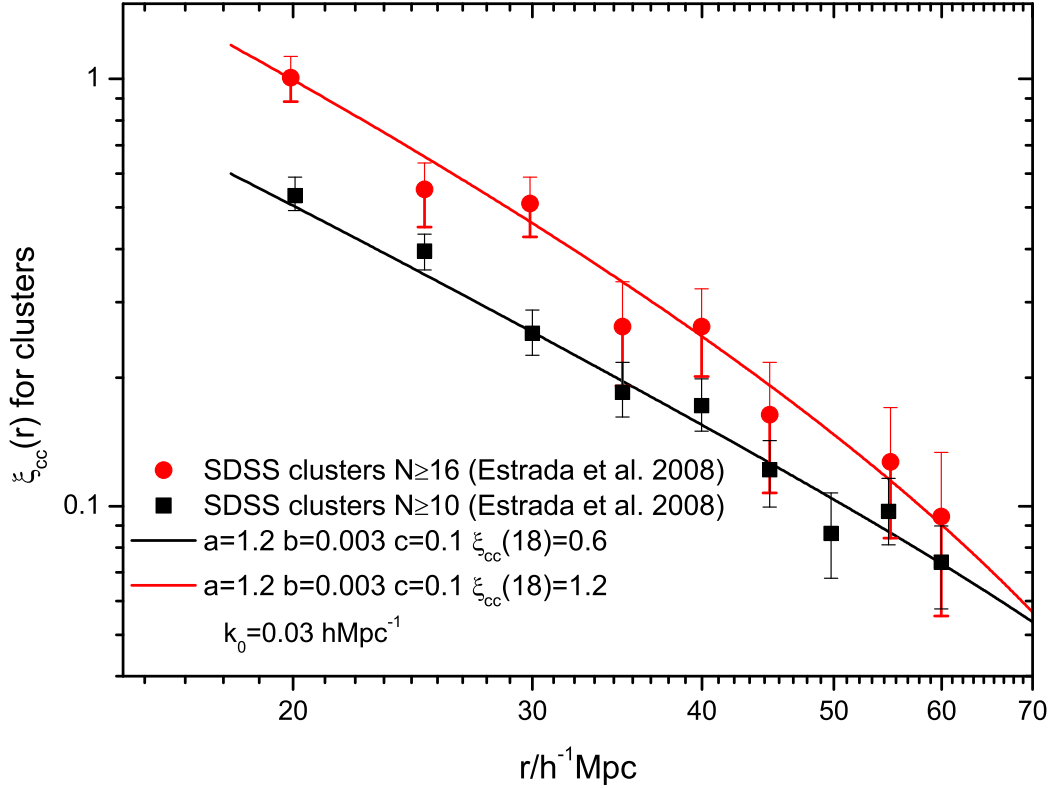}
	\caption{The solution $\xi_{cc}(r)$ matches
		the cluster data of SDSS  with two types of richness (\cite{Estradaetal(2009)}).
		Notice that  $(a,b,c)$ are the same as for galaxies,
		but $k_0=0.03 \, h$Mpc$^{-1}$ taken for clusters is smaller than that for galaxies.}
	\label{BahcallFig}
\end{figure}
In Fig. \ref{BahcallFig}
two solutions $\xi_{cc}(r)$ with different amplitudes are given,
and are compare with two sets of data with richness $N> 10$ and  $N> 16$
from the SDSS  (\cite{Estradaetal(2009)}).
Interpreted by the field equation (\ref{eqfinal}),
the $N> 16$  clusters have a greater $m$
than the $N> 10$ clusters.
The solutions match the data  available
on the whole range $r=(18\sim 60) \, h^{-1}$Mpc,
and there is no small-scale insufficiency of correlation
that occurred for the galaxy case.
This means that the order of $(\delta\psi)^2$ in  perturbations  is accurate enough
to account for the correlation of clusters.
Since $k_0=0.03 \, h$Mpc$^{-1}$ for clusters and $k_0=0.055 \, h$Mpc$^{-1}$ for galaxies,
it can be inferred that
the mean density $\rho_0$ involved in this cluster survey
is lower by $(0.03/0.055)^2\sim 0.3$
than those in the galaxy case.

\begin{figure}
	\includegraphics[width=14.0cm, angle=0]{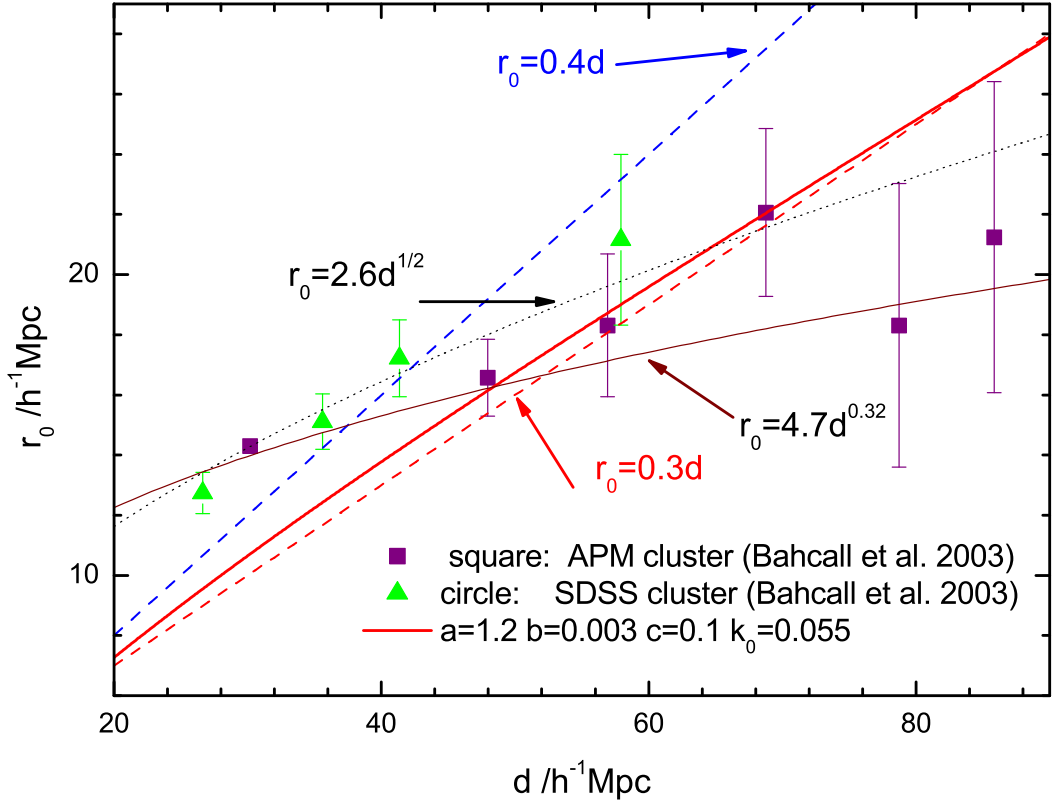}
	\caption{The solution $\xi_{cc}(r)$ with $k_0=0.03 \, h$Mpc$^{-1}$
		gives the scaling  $r_0\simeq 0.4d$.
		But with a greater $k_0=0.055 \, h$Mpc$^{-1}$,
		$\xi_{cc}$ would give a flatter scaling $r_0\simeq 0.3d$,
		which seems to fit the data of APM clusters better (\cite{Bahcalletal(2003)}).}
	\label{figscaling}
\end{figure}

Observations show that
the cluster correlation scale increases
with the mean spatial separation  between clusters
(\cite{SzalaySchramm(1985),BahcallWest(1992),Bahcall(1996),
Croftetal(1997),Gonzalezetal(2002)}).
For a power-law $\xi_{cc} =(r_0/r)^{1.8}$ fitting,
the data indicates a  ``correlation length"
\be  \label{scaling}
r_0 \simeq 0.4d_i,
\ee
where $d_i=n_i^{-1/3}$ and $n_i$
is the mean number density of clusters of type $i$.
For SDSS, the scaling can be also fitted by
$r_0 \simeq 2.6 d_i\,^{1/2} $ (\cite{Bahcalletal(2003)}),
and for the 2dF galaxy groups $r_0\simeq 4.7 d_i\,^{0.32}$ (\cite{Zandivarezetal(2003)}).
This kind of  $r_0-d_i$ empirical scaling   has been a theoretical
challenge (\cite{Bahcall(1996)}),
and was thought to be either caused
by a fractal phenomenon (\cite{SzalaySchramm(1985)}),
or by the statistics of rare peak events (\cite{Kaiser(1984)}).
Interpreted by our theory,
the scaling behavior is completely contained  in the solution $\xi_{cc}(k_0 r)$,
where the characteristic wavenumber $k_0  \propto \rho_0^{1/2} \propto d^{-3/2}$
appears together with $r$ in  the variable of the function $\xi_{cc}$.
To comply with the empirical power-law,
we take the theoretical ``correlation length" as $r_0(d) \propto \xi_{cc}^{1/1.7}$,
where $\xi_{cc}$ is the solution   depending  on  $d$.
Fig.\ref{figscaling} shows that
the solution $\xi_{cc}$ with $k_0=0.03 \, h$Mpc$^{-1}$
gives the scaling  $r_0(d)\simeq 0.4d$,
agreeing well with the observation (\cite{Bahcall(1996)}).
If  a greater $k_0=0.055 \, h$Mpc$^{-1}$ is taken,
the solution $\xi_{cc}$ would yield a flatter scaling  $r_0(d)\simeq 0.3d$,
which fits the data of APM clusters better (\cite{Bahcalletal(2003)}).
Our solution $\xi_{cc}(k_0r)$ tells that a higher background density $\rho_0$
corresponds to a flatter slope of the scaling  $r_0(d)$,
and the scaling is naturally explained.

Extended to very large scales,
the observed $\xi_{cc}(r)$ exhibits
a pattern of periodic oscillations
with a characteristic wavelength $\sim 120 \, h^{-1}$Mpc
 (\cite{Einastoetal(1997a),Einastoetal(1997b)}).
It  was originally
found in the galaxy distribution in narrow pencil beam
surveys (\cite{Broadhurstetal(1990)}),
also occurred in the correlation function of galaxies (\cite{Tuckeretal(1997)}),
and of quasars (\cite{Yahataetal(2005)}).
There have been various interpretations.
Our solution  $\xi(r)$ with small values $(a,b,c)$ exhibits periodic oscillations
with a damped amplitude at increasing $r$ (\cite{Zhang(2007)}).
Although the data from samples of a cylindrical volume
show  large amplitude at $r\sim 400- 600 \, h^{-1}$Mpc,
these high amplitude would be damped  in a full 3-dim sample (\cite{Einastoetal(2002)}).

\section{3-point Correlation  Function in Gaussian Approximation}
\label{Sect:3PTCF}

It is also interesting to consider the 3-point correlation
function $G^{(3)}(\bf r,r',r'')$
in the Gaussian approximation in our theory.
As given in Appendix \ref{App:3PTCF},
the field  equation of $G^{(3)}$ in  Gaussian approximation is
\ba \label{eqG3p}
&& \nabla ^2_{r}   G^{(3)} ( {\bf r,r',r'' } )
+ 2 k_J ^2   \psi_0  G^{(3)} ( {\bf r,r',r'' } )
- \frac{2}{\psi_0} \nabla  G^{(2)} ( {\bf r,r''}) \cdot \nabla  G^{(2)} ( {\bf r,r'} ) \nn \\
&&+  2k_J ^2   G^{(2)} ( {\bf r,r'} ) G^{(2)} ( {\bf r,r''} )
+ \frac{2}{\alpha}    \psi_0 \delta^{(3)}( {\bf r-r''} ) G^{(2)} ({\bf r,r'})
+ \frac{2}{\alpha}  \psi_0 \delta^{(3)}({\bf r-r'})  G^{(2)} ({\bf r,r''} )  =0 , \nn \\
\ea
To look for its solution, let $G^{(3)}$
be of the  form of the Kirkwood-Groth-Peebles  anzats (\ref{Ansatz})  with
\be
Q= 1/\psi_0.
\ee
In fact,   $Q= 1$
since $\psi_0 =1$ by  $\langle \rho\rangle =\rho_0$.
Using  Eq. (\ref{Helmoltz})
for the 2-point function $G^{(2)} $ at Gaussian approximation
and  the property of  $\delta$-function,
the field equation (\ref{eqG3p})  is satisfied automatically
(See Appendix \ref{App:3PTCF}).
Thus,
at the Gaussian approximation  of our theory,
the analytical solutions are
\be\label{3ptsol}
G^{(3)}(\textbf{r}_1,\textbf{r}_2,\textbf{r}_3)
= G^{(2)}(r_{12} ) G^{(2)}(r_{23} )
+G^{(2)}(r_{23} ) G^{(2)}(r_{31} )
+G^{(2)}(r_{31} ) G^{(2)}(r_{12} ),
\ee
where the Gaussian 2-point correlation function (\cite{Zhang(2007)})
\be\label{2ptsol}
G^{(2)}({\bf r})
= A_1 \frac{Gm}{c^2_s}\frac{\cos(k_0 r)}{r}
+ A_2 \frac{Gm}{c^2_s}\frac{ \sin (k_0 \, r)}{r},
\ee
with  the coefficients satisfying $A_1+A_2=1$.
This result proves  that
the Kirkwood-Groth-Peebles ansatz (\ref{Ansatz})
with $Q= 1$ holds exactly  in Gaussian approximation.
As for the  nonlinear field equation of $G^{(3)}$ beyond Gaussian approximation,
it will be much  more involved,
and will be studied  in future.

\section{Conclusions and Discussions }
\label{Sect:ConDisc}

We have presented a field theory of density fluctuations of
a Newtonian self-gravitating system,
derived  the nonlinear field equation  of the correlation function
of second order of perturbations,
and applied it to  systems galaxies and of clusters in the Universe.

As the starting point,
we have obtained the field equation (\ref{masseq}) of the mass density field $\psi$
under the condition of  hydrostatic equilibrium.
It suits better for studying the mass distribution
than the Lane-Emden equation of gravitational potential.
In dealing with the high nonlinearity,
the mass density field is expanded as $\psi=\psi_0+\delta\psi$,
where the mean density $\psi_0$ is a constant for the background of universe
and the fluctuation is kept to the order $(\delta\psi)^2$ in this paper.
This approach can also apply to other finite self-gravitating systems,
in which a given background density $\psi_0$ varies in space.
As the main result, the field equation (\ref{eqfinal})
of 2-point correlation function $G^{(2)}$ of density fluctuation has been derived,
whereby the Kirkwood-Groth-Peebles ansatz is adopted to cut off the hierarchy
and renormalization is performed.
As for the 3-point correlation $G^{(3)}$,
it is very revealing  to find that
its  field  equation  at Gaussian approximation
is automatically satisfied
when  the Kirkwood-Groth-Peebles ansatz  is used   with $Q= 1$.
Thus  the  ansatz holds as an exactly relation  between  $G^{(2)}$  and  $G^{(3)}$
at Gaussian level in our theory.
The equation of  $G^{(2)}$
is Helmholtz-like and nonlinear up to order $(G^{(2)})^2$,
with three parameters $(a,b,c)$ representing  nonlinear effects
beyond the Gaussian approximation.
Notably,
the Jeans wavelength $\lambda_0$ occurs as the only scale,
and the mass $m$ appears as the magnitude of the source.
The result simultaneously explains
several seemingly unrelated features of large scale structure of the Universe,
such as
the profile similarity of $\xi_{cc}$ for clusters
to $\xi_{gg}$  for  galaxies,
the differences in amplitude and in correlation length
of $\xi_{cc}$ and $\xi_{gg}$,
the  $r_0-d$ scaling,
and the pattern of periodic oscillations in $\xi_{cc}$
with a wavelength $\lambda_0 \sim 120 \, h^{-1}$Mpc.
With the same set of $(a,b,c)$,
the  solution $\xi_{gg}$  agrees with  observations of galaxies
over a range $(1\sim 50)\, h^{-1}$Mpc,
and the solution  $\xi_{cc}$ of larger $m$ and $\lambda_0$
matches observations  of clusters over the whole range $(4\sim 100) \, h^{-1}$Mpc.
Thus,  our theory sheds light on the understanding of
the large scale structure of Universe.

There are several  possible improvements of the present work.
To improve the correlation at small scales $r \leq 1$ Mpc,
higher order  fluctuations   are needed,
and this can be carried out systematically by perturbation.
To include the evolution effects,
the field equation of correlation should be extended to the case of cosmic expansion.
Finally, our formulation  of perturbation  can be systematically used
to derive the nonlinear field equations of $G^{(3)}$ beyond Gaussian approximation, etc.
These would need more studies.

\section*{Acknowledgements}

Y. Zhang is supported by
NSFC Grant No. 11421303, 11675165, 11633001
SRFDP, and CAS, the Strategic Priority Research Program
``The Emergence of Cosmological Structures"
of the Chinese Academy of Sciences, Grant No. XDB09000000.

\appendix
\numberwithin{equation}{section}

\section[]{Grand Partition Function as a Path Integral}
\label{App:GPF}

From the identity
\begin{align}
&\exp{ \left[ \frac{1}{2}m^2V \right]}
=\frac{1}{ \sqrt{2\pi V}}
\int_{-\infty}^{\infty}dx
\exp{ \left[ -\frac{1}{2V} x^2 +m x  \right]},
\,\,\, {\rm with}\,\,V>0,
\end{align}
one can extend to the Stratonovich-Hubbard identity (\cite{Stratonovich(1957),Hubbard(1959)}):
\ba
\exp{ \left[ \frac{1}{2}\sum_{i,j}^N  m_i V_{ij} m_j \right]}
=\frac{1}{ \sqrt{\det(2\pi V)}} \prod_i^N \int_{-\infty}^{\infty}dx_i
\exp{ \left[ -\frac{1}{2}\sum_{i,j}^N  x_iV^{-1} _{ij} x_j
	+\sum_i ^N  x_i m_i  \right]},
\ea
where $(V_{ij})$ is a symmetric matrix with positive eigenvalues.
This can be further extended to the continuous case.
Let $V( r)$ be a long range attractive potential, and its inverse
$K$  as a kernel is defined  by
\be
\int d^3r   K( {\bf r}_1-{\bf r} )  V({\bf r}-{\bf r}_2)
=\delta^{(3)}({\bf r}_1- {\bf r}_2).
\ee
Then the Stratonovich-Hubbard identity in this continuous case is (\cite{Zinn-Justin(1996)})
\be
\exp{ \left[ \frac{1}{2}T^{-1} \sum_{i,j} V(r_i-r_j) \right]}
=
\mathcal{N} \int_{-\infty}^{\infty}D \phi
\exp{ \left[- \frac{1}{2} T
	\int d^3r_1 d^3r_2 \phi(r_1)K(r_1-r_2) \phi(r_2)
	+\sum_i ^N \phi(r_i)  \right]},
\ee
where $T$ is a constant,
and the numerical factor $\mathcal{N}\propto 1/\sqrt{ \det V}$
is a multiplicative factor to the grand partition function $Z$,
irrelevant to the ensemble averages of physical quantities,
and  can be dropped.
For a formally stricter treatment,
a hard core of radius $r_c$, say the size of a typical galaxy,
should have been introduced at the center of $V(r)$
so that there would be a cutoff of lower limit of integration to avoid
the divergence.
But this divergence will only occur in $\mathcal{N}$
and is dropped off eventually.

The interesting case is the  potential
$V({\bf r}_1- {\bf r}_2) = \frac{1}{|{\bf r}_1- {\bf r}_2|}$.
By
\be
\nabla^2 \frac{1}{|{\bf r}_1- {\bf r}_2|}
=-4\pi \delta^{(3)}({\bf r}_1- {\bf r}_2),
\ee
the kernel is
$ K( {\bf r}_1-{\bf r} )
= - \frac{1}{4\pi}\delta^{(3)}( {\bf r}_1-{\bf r})\nabla^{2}$.
Integrating by parts, one has
\be
\int d^3r_1 d^3r_2 \phi(r_1)K(r_1-r_2) \phi(r_2)
= \frac{1}{4\pi}\int d^3r (\nabla \phi)^2,
\ee
so that  (\cite{deVegaetal(1996b)})
\begin{align}
\exp{ [ \frac{1}{2} T^{-1} \sum_{i,j}^N
	\frac{Gm^{2} }{|{\bf r}_i- {\bf r}_j|} ]}
=\int_{-\infty}^{\infty}D \phi
\exp{ [- \frac{1}{2}  \alpha
	\int d^{3}r (\nabla \phi)^2
	+\sum_i ^N \phi(\bf r_i) ]},
\label{partialintegral}
\end{align}
where $\alpha \equiv T/4\pi  Gm^2 $.
The term $\sum_i \phi(\bf r_i)$ in Eq. (\ref{partialintegral})
is a sum of interactions of the field $\phi$
with the  point mass at $\bf r_i$
(and could also be written as an integration
$\sum_i \phi({\bf r_i})= \int d^3r \phi({\bf r})n(\bf r)$
where $n(\bf r)$ is the number density of particles).

We use the above result to write
the grand partition function  $Z$  in Eq. (\ref{Z1})
as a path integral in Eq. (\ref{Zphi}).
The kinetic energy term in $e^{-H/T}$ in Eq. (\ref{Z1}) after integrating over
the momentum $ d^3p_i$ gives
\be
\int \frac{d^3p_i }{(2\pi)^3}
\exp[-p_i^2/2mT]=(\frac{mT}{2\pi})^{3/2}.
\ee
The potential term in $e^{-H/T}$ is given by Eq. (\ref{partialintegral}),
in which only $\sum_i ^N \phi(r_i)$ involves
integration over the coordinates
\be
\int \prod_{i=1}^N d^3r_i \exp{ \sum_i ^N \phi(r_i)}=
\left[\int d^3r  \exp{  \phi(r)} \right]^N.
\ee
Thus one has the grand partition function
\begin{align}
\label{ZA}
&Z=\sum_{N=0}^\infty \frac{1}{N!}[z(\frac{mT}{2\pi })^{3/2} ]^N
\cdot\int_{-\infty}^{\infty}D \phi
\exp{ \left[- \frac{1}{2}
	\alpha \int d^3r (\nabla \phi)^2 \right]}
\left[\int d^3r  \exp{  \phi(r)} \right]^N  \nonumber \\
&=\int_{-\infty}^{\infty}D \phi
\exp{ \left[- \frac{1}{2}
	\alpha \int d^3r (\nabla \phi)^2
	+ z(\frac{m T }{2\pi})^{3/2}\int d^3r e^{\phi(r)} \right]}.
\end{align}
Using the fugacity $z=(2\pi/mT)^{3/2}n_0$
for a dilute gas of the mean number density $n_0$,
one obtains $Z$ as in  Eq. (\ref{Zphi}).
The derivation can be also found in Ref. (\cite{deVegaetal(1996b)}).

\section{Derivation of  Field Equation	and  Renormalization}
\label{App:FEandRenorm}

We present the derivation of the field equation of
the 2-point correlation function $G^{(2)}(\textbf{r}-\textbf{r}^{\prime})$.
The technique involved is the functional differentiation of
the generating functional $Z[J]$ in Eq. (\ref{ZJ})
with respect to the external source $J$.
The method was initiated by Schwinger,
and has been commonly adopted in field theory of particle physics and
of condensed matter physics  (\cite{Goldenfeld(1992)}).
We start with the ensemble average of Eq. (\ref{field_eq})
of the mass density field in the presence of  $J$,
\be \label{main0}
\langle \, \nabla^{2}\psi({\bf r} )
-\frac{(\nabla\psi({\bf r} ))^{2}}{\psi({\bf r} )}
+k_{J}^{2}\psi({\bf r} )^{2}+J\psi({\bf r} )^{2} \, \rangle_J  =0,
\ee
differentiate it with respect to $J$
\begin{equation} \label{main}
\frac{\delta}{\delta J({\bf r} ^{\prime})}
\langle  \, \nabla^{2}\psi({\bf r} )
-\frac{(\nabla\psi({\bf r} ))^{2}}{\psi({\bf r} )}
+k_{J}^{2}\psi({\bf r} )^{2}+J({\bf r})\psi({\bf r} )^{2}  \, \rangle_J=0,
\end{equation}
and set $J=0$, and shall  end up with
the field equation for $G^{(2)}({\bf r}-{\bf r}')$.
In the following we deal with each term of Eq. (\ref{main}).

The first term of Eq. (\ref{main}) gives
\be \label{1b2}
\langle\nabla^{2}\psi({\bf r})\rangle _J
= \nabla^{2}\langle\psi( {\bf r} )\rangle_J .
\ee
Changing the ordering of $\frac{\delta}{\alpha \delta J({\bf r} ^{\prime})} $
and $\nabla^{2}$,
using the definition (\ref{2ptGreen}),
we obtain
\be \label{nablaG}
\nabla^{2}\left(\frac{\delta}{\alpha \delta J(\bf{r}^{\prime})}
\langle\psi({\bf r})\rangle_J\right)|_{J=0}
= \nabla^{2}G^{(2)}(\bf{r}-\bf{r}^{\prime}).
\ee
For the remaining three terms of Eq. (\ref{main}),
we firstly work for the Gaussian approximation (\cite{Zhang(2007)}),
i.e,
the lowest order of approximation that includes the fluctuation $\delta\psi$
beyond the mean-field approximation.
Split the field $\psi $  into the  averaged and perturbation parts:
\[
\psi= \langle  \psi \rangle_J  +\delta\psi.
\]
Take
\be \label{1/psi}
\frac{1}{\psi}=\frac{1}{ \langle  \psi \rangle_J  +\delta\psi}
\simeq \frac{1}{ \langle  \psi \rangle_J }  -\frac{\delta\psi}{ \langle  \psi \rangle_J ^2 },
\ee
where terms   $( \delta\psi  )^2$
and higher have been  dropped,
and
\be
\langle\frac{(\nabla\psi)^{2}}{\psi}\rangle_J
\simeq \frac{ (\nabla \langle\psi\rangle_J)^{2}}{\langle\psi\rangle_J},
\ee
in the Gaussian approximation.
Applying $\frac{\delta}{\delta J({\bf r} ^{\prime})}$ to this term,
and using
$\langle\psi(\textbf{r})\rangle|_{J=0} =\psi_{0}=1$
and $\nabla \psi_0=0$,
one finds that
the second term yields a vanishing contribution
to Eq. (\ref{main}).
The third and fourth terms involve $\psi^2$.
By
\be \label{psisqq}
\langle \psi^2 \rangle_J
=  \langle  \psi \rangle_J^2  +2 \langle  \psi \rangle_J \langle \delta\psi \rangle_J
+  \langle(\delta\psi)^2\rangle_J
\simeq \langle  \psi \rangle_J^2  ,
\ee
where $\langle \delta\psi \rangle_J=0$ is used,
and $\langle(\delta\psi)^2\rangle_J $ is dropped as a high order term.
By  Eq. (\ref{2ptGreen}) and
\be \label{deltaJ}
\frac{\delta J({\bf r})}{\delta J({\bf r}')}
=\delta^{(3)}({\bf r}-{\bf r}^{\prime}),
\ee
Eq. (\ref{main}) in the Gaussian approximation yields
the Helmholtz equation with a point source
\be \label{Helmoltz}
\nabla^{2}G^{(2)}({\bf r} )+k_{0}^{2}G^{(2)}({\bf r} )
=-\frac{1}{\alpha } \delta^{(3)}({\bf r} ),
\ee
where $k_0=\sqrt{2}k_J$ is the characteristic wavenumber.
The term $+k_0^{2}G^{(2)}$
has a plus sign because  gravity is attractive.
The Gaussian solution  is of the form:
\be \label{GGaussian}
G^{(2)}({\bf r}) \propto \frac{Gm}{c^2_s}\frac{\cos(k_0 r)}{r},\,\,\,\,\,\,
\frac{Gm}{c^2_s}\frac{ \sin (k_0 \, r)}{r},
\ee
subject to certain boundary condition in specific applications.
A general solution will be a linear sum of  the two.
The choice $\cos(k_0 r)/r$   confronts with the data
of galaxy surveys better (\cite{Zhang(2007)}).
From Eq. (\ref{GGaussian}), we find an important property
that the amplitude is proportional to the mass $m$:
\be \label{proptom}
G^{(2)} \propto m.
\ee
By Fourier transform
\be
G^{(2)}({\bf r})=\iiint  P(k)e^{i\bf k\cdot r}d^3 {\bf k }
\ee
the power spectrum in the Gaussian approximation is
\be \label{Pkgaussian}
P(k)=\frac{1}{2n_0}\frac{1}{(k/k_0)^2-1},
\ee
telling that $P(k)$ is higher for galactic
objects with a lower spatial number density $n_0$.
In this paper, we take $k_0 \simeq 0.055\, h$Mpc$ ^{-1}$
for the  system of galaxies.
Eq. (\ref{Pkgaussian}) is also arrived at in Ref. in another
context (\cite{Chavanis(2006)}).
In literature on large scale structure (\cite{Peacock(1999)}),
a dimensionless spectrum
\be
\Delta^2 (k)  \equiv V\frac{2}{\pi^2} k^3 P(k).
\ee
with $V$  being a normalization volume
is often used.
As we have demonstrated  in Ref.(\cite{Zhang(2007)}),
for large scales $k = (0.05 \sim 0.3)\, h$Mpc$ ^{-1}$,
Eq. (\ref{Pkgaussian}) predicts  $P(k) \propto k^{-2.2}$,
i.e, $ \Delta^2 (k)  \propto k^{0.8}$ ,
agreeing   qualitatively with
galaxy  surveys (\cite{Peacock(1999),Carrettietal(2001),Dodelsonetal(2002),
Tegmarketal(2004),Coleetal(2005)}).
But on small scales,
one should
and use  $G^{(2)}$ beyond Gaussian approximation,
which   gives  $P(k) \propto k^{-1.6}$
for   scales $k = (0.05 \sim 0.7)\, h$Mpc$ ^{-1}$ (\cite{ZhangChen(2015)}).
This improves by including more nonlinear clustering.

We just mention that
the similar form of Eq. (\ref{Helmoltz})
also occurs in the the Gaussian approximation of the Landau-Ginzburg theory
of phase transition (\cite{Goldenfeld(1992),Binneyetal(1992)}),
where $G^{(2)}(\bf r)$ is also called the bare propagator.
However, in Landau-Ginzburg theory,
the corresponding term $-\mu^2G^{(2)}(\bf r)$, in place of $+k_0^22G^{(2)}(\bf r)$,
has a negative sign,
the solution is $G^{(2)}({\bf r}) \propto e^{-\mu r}/r$
with $1/\mu$ being the correlation length.
At the critical point of phase transition, $\mu\rightarrow 0$,
$G^{(2)}({\bf r}) \propto 1/r$,
the correlation becomes  long-range.
In contrast, in our case,
gravity is a long range attractive interaction,
the self-gravitating system is long-range correlated,
as evidenced by
the fact that  $G^{(2)}({\bf r}) $ in Eq. (\ref{GGaussian})
has $\cos(k_0 r)$ and $\sin(k_0 r)$,
instead of the exponential decay $e^{-\mu r}$.
In this sense, the self-gravitating system is always
at the critical point of phase transition (\cite{Saslaw+2000}).

In the field theories,
one considers a physical field $\psi$.
The expectation value $\langle \psi \rangle$
formally contains all information of fluctuation of the system,
only in the sense that one has to perturb
the  system with an external source $J$
(a test magnetic field, or a test mass, etc)
and  observes the response of the system.
This procedure is just carried out by
applying
$\frac{\delta}{  \delta J({\bf r}')} $ on $\langle\psi({\bf r})\rangle_J|_{J=0}$
to get the response function
$G^{(2)}({\bf r},{\bf r}')=\frac{\delta}{\alpha \delta J({\bf r}')} \langle\psi({\bf r})\rangle_J|_{J=0}$.
which is just the 2-point correlation function  $G^{(2)}({\bf r},{\bf r}')$,
telling us how
$\langle \psi \rangle$ of system changes under the applied, external $J$.
In condensed matter  $G^{(2)}({\bf r},{\bf r}')$ stands for the susceptibility
of the system.
In our case, $ G^{(2)}({\bf r},{\bf r}')$ is essentially
the probability of finding  a galaxy at $\bf r'$ above the average,
given a galaxy at $\bf r$.
For a Gaussian system, $G^{(2)} $
contains all the statistical, dynamical information of the system.
But for  the self-gravity system,
higher order correlation functions $G^{(3)}$,  $G^{(4)}$, etc,
are need in order to have more statistical and dynamical information of the system.

Now beyond the Gaussian approximation,
we shall include high order terms of the fluctuation $\delta \psi$
in  Eq. (\ref{main}).
The final nonlinear equation of $G^{(2)}({\bf r}) $
will include  terms up to  $(\, G^{(2)} \,)^2$ in this paper.

The third term of Eq. (\ref{main}) now is
\begin{align}
\langle \psi^{2}\rangle_J
=\langle\psi \rangle_J^{2}+\langle\delta\psi\delta\psi\rangle_J,
\end{align}
where   $(\delta\psi)^2 $ is kept,
in contrast to Eq. (\ref{psisqq}) of the Gaussian approximation.
Applying $\frac{\delta}{\alpha \delta J(\bf{r}')}$ to the above yields
\begin{align} \label{3b2}
& \frac{\delta}{\alpha\delta J(\textbf{r}^{\prime})}
( \langle\psi \rangle_J^{2}+\langle\delta\psi\delta\psi\rangle_J )|_{J=0}
=2\psi_{0}   G^{(2)}(\textbf{r}-\textbf{r}^{\prime})
+ G^{(3)}(\textbf{r},\textbf{r},\textbf{r}^{\prime}),
\end{align}
where  the 3-point correlation function
$ G^{(3)}({ \bf r},{\bf r},{\bf r}')=
\frac{\delta}{\alpha \delta J({\bf r}')}\langle\delta\psi\delta\psi\rangle|_{J=0}$
is used.
In our previous treatment  (\cite{ZhangMiao(2009)}),
$G^{(3)}$ in the above was dropped as a high-order term.

The fourth term of Eq. (\ref{main}) is
\be
\langle J\psi^{2}\rangle_J= J \langle \psi^{2}\rangle_J
=J\langle\psi \rangle_J ^{2}+J\langle\delta\psi\delta\psi\rangle_J,
\ee
where $(\delta\psi )^2 $ is also kept.
Applying $\frac{\delta}{\alpha \delta J(\bf{r}')}$ to the above and using  Eq. (\ref{deltaJ}),
one has
\be \label{4b2}
\frac{\delta}{\alpha \delta J(\textbf{r}^{\prime})}\langle J\psi^{2}\rangle_J |_{J=0}
=\frac{1}{\alpha}\big( \, \psi_{0}^{2}+G^{(2)}(0)  \, \big)\delta^{(3)}(\bf{r}-\bf{r}^{\prime}),
\ee
where
$G^{(2)}(0)=\langle\delta\psi\delta\psi\rangle
=\lim_{r'\rightarrow r}G^{(2)}({\bf r}-{\bf r}')$.
For the system of galaxies,
the definition of $G^{(2)}(\bf r)$ applies
only for $r> r_c$ with $r_c$ being the galaxy size.
The occurrence of the quantity $G^{(2)}(0)$
is inevitable
when high order terms of $\delta\psi$  are included beyond the Gaussian approximation.
This is common in  field theory  with interactions,
both in particle physics and condensed matter physics.
In the former case, the analogue of $G^{(2)}(0)$ is divergent,
and, in the latter, a cutoff is introduced for $|{\bf r}-{\bf r}'|\ge r_c$,
and $G^{(2)}(0)$ is finite.
For example, in our case,
it could be expressed as an integration over the momentum
\be
G^{(2)}(0)=\lim_{r\rightarrow r'}
\int d^3k \frac{e^{-i{\bf k\cdot({\bf r}-{\bf r}') }}}{k^2-k_0^2}
=\int d^3k \frac{1}{k^2-k_0^2}
\ee
of the ``bare" propagator $1/(k^2-k_0^2)$ of the Gaussian approximation.
Later $\nabla G^{(2)}(0)$ and $\nabla^2 G^{(2)}(0)$
will occur and have the similar expressions, correspondingly.
These three quantities are undetermined
and can be handled by a  renormalization  procedure,
by which these quantities are eventually absorbed into  physical quantities,
such as   mass,   field amplitude,   coupling constant, etc,
depending on the specific field theory  (\cite{Binneyetal(1992)}).
In this paper, similarly,
we shall use renormalization to
absorb $ G^{(2)}(0)$,  $\nabla G^{(2)}(0)$, and $\nabla^2 G^{(2)}(0)$.

The second term of Eq. (\ref{main}) contains a factor $1/ \psi$.
Instead of Eq. (\ref{1/psi}) in the Gaussian approximation,
we now keep up to the order of  $(\delta\psi)^2$:
\be \label{expand}
\frac{1}{\psi}=\frac{1}{ \langle  \psi \rangle  +\delta\psi}
\simeq
\frac{1}{ \langle  \psi \rangle }
\left(1-\frac{\delta\psi}{ \langle  \psi \rangle }+
(\frac{\delta\psi}{\langle  \psi \rangle })^{2} \right).
\ee
Here and in the following the subscript ``$J$" in $\langle  \psi \rangle_J$
is skipped  temporarily for simplicity.
This perturbation is a good approximation for $\delta\psi/\langle\psi\rangle \ll 1$.
At small scales where  $\delta\psi$ is large,
higher nonlinearity would  be needed   than this order of perturbations.
Using  Eq. (\ref{expand}), one has
\begin{align}\label{2}
\langle\frac{(\nabla\psi)^{2}}{\psi}\rangle
&=\frac{1}{ \langle \psi \rangle}\langle\left(1-\frac{\delta\psi}{\langle\psi \rangle}
+(\frac{\delta\psi}{\langle  \psi \rangle })^{2} \right)
(\nabla  \langle\psi\rangle +\nabla\delta\psi)^{2} \rangle\nonumber\\
&\simeq \frac{(\nabla\langle\psi\rangle)^{2}}{\langle\psi\rangle}
+\frac{\langle(\nabla\delta\psi)^{2}\rangle}{\langle\psi\rangle}
-\frac{2\nabla \langle  \psi \rangle}{ \langle  \psi \rangle^{2}}
\cdot \langle\delta\psi\nabla\delta\psi\rangle
+\frac{(\nabla \langle \psi\rangle )^{2}}{\langle \psi\rangle^{3}}
\langle(\delta\psi)^{2}\rangle,
\end{align}
where  $(\delta\psi)^3$ and higher are dropped.
Note that,
$(\delta\psi)^2$
in the expansion of $\langle\frac{(\nabla\psi)^{2}}{\psi}\rangle$
is always  accompanied by  a factor
$\delta\langle \psi \rangle  / \delta J \sim G^{(2)}$ in deriving the equation.
Thus,    the final equation of $G^{(2)}$
actually contains the fluctuation  up to the order  $( \delta\psi)^4$
in this  paper.

(\ref{2}) contains four sub-terms.
The first and second terms of (\ref{2}),
$\frac{(\nabla  \langle  \psi \rangle)^{2}}{ \langle  \psi \rangle}
+\frac{\langle(\nabla\delta\psi)^{2}\rangle}{ \langle  \psi \rangle}$,
will be treated together.
In our previous treatment (\cite{ZhangMiao(2009)}),
$\langle(\nabla\delta \psi)^{2}\rangle\rightarrow\nabla^{2}\langle(\delta\psi)^{2}\rangle$
was improperly taken.
Now we treat it in the following.
By the field equation (\ref{field_eq}), one has
\be\label{(1)}
(\nabla\psi)^{2}=\psi\nabla^{2}\psi+(k_{J}^{2}+J)\psi^{3},
\ee
and there is an identity:
\be
(\nabla\psi)^{2}=\nabla\cdot(\psi\nabla\psi)-\psi\nabla^{2}\psi\label{(2)}
=\frac{1}{2}\nabla^2 \psi^2  -\psi\nabla^{2}\psi.
\ee
Adding Eq. (\ref{(1)}) and Eq. (\ref{(2)}) together and taking the ensemble average,
one has
\be
\langle(\nabla\psi)^{2}\rangle=\frac{1}{4}\langle\nabla^{2}\psi^{2}\rangle
+\frac{1}{2}(k_{J}^{2}+J)\langle\psi^{3}\rangle.
\ee
Substituting $\psi=\langle\psi\rangle+\delta\psi$ into the above leads to
\begin{align}
(\nabla \langle\psi\rangle)^{2}+\langle(\nabla\delta\psi)^{2}\rangle
=\frac{1}{4}\nabla^{2}\langle\psi \rangle^{2}+\frac{1}{4}\nabla^{2}\langle\delta\psi\delta\psi\rangle
+\frac{1}{2}(k_{J}^{2}+J) \langle  \psi \rangle^{3}
+\frac{3}{2}(k_{J}^{2}+J) \langle  \psi \rangle\langle\delta\psi\delta\psi\rangle,
\end{align}
where the higher order term $(\delta\psi)^{3}$ is dropped.
Thus, the first and second terms of (\ref{2}) are given by
\begin{align} \label{sub12}
\frac{(\nabla  \langle  \psi \rangle)^{2}}{ \langle  \psi \rangle}
+\frac{\langle(\nabla\delta\psi)^{2}\rangle}{ \langle  \psi \rangle}
=\frac{1}{4} \frac{\nabla^{2}\langle  \psi \rangle^{2}}{\langle  \psi \rangle}
+\frac{1}{4}\frac{\nabla^{2}\langle\delta\psi\delta\psi\rangle}{\langle  \psi \rangle}
+\frac{1}{2}(k_{J}^{2}+J)\langle  \psi \rangle ^{2}
+\frac{3}{2}(k_{J}^{2}+J) \langle\delta\psi\delta\psi\rangle.
\end{align}
Applying  $\frac{\delta}{\delta J({\bf r} ^{\prime})} $ to Eq. (\ref{sub12})
and setting $J=0$ yields the contribution of
the first and second terms of Eq. (\ref{2}):
\begin{align}  \label{2.a+2.b}
&\frac{\delta}{\alpha  \delta J(\textbf{r}^{\prime})}
\left(\frac{(\nabla  \langle  \psi \rangle)^{2}}{ \langle  \psi \rangle}
+\frac{\langle(\nabla\delta\psi)^{2}\rangle}{ \langle  \psi \rangle} \right)|_{J=0}\nonumber\\
&= \frac{1}{2}\nabla^{2} G^{(2)}(\textbf{r}-\textbf{r}^{\prime})
+k_{J}^{2}\psi_{0}G^{(2)}(\textbf{r}-\textbf{r}^{\prime})
-\frac{1 }{4\psi_{0}^{2}} \nabla^{2}G^{(2)}(0) G^{(2)}(\textbf{r}-\textbf{r}^{\prime}) \nonumber\\
&+(\frac{1 }{4\psi_{0}}\nabla^{2}
+\frac{3}{2} k_{J}^{2} )G^{(3)}(\textbf{r},\textbf{r},\textbf{r}^{\prime})
+\frac{1}{2\alpha }\big( \, \psi_{0}^{2}+3G^{(2)}(0)  \, \big)\delta^{(3)}(\textbf{r}-\textbf{r}^{\prime}).
\end{align}
where
$\nabla^{2} G^{(2)}(0)=\nabla^{2}\langle\delta\psi\delta\psi\rangle
=\lim_{r\rightarrow 0}\nabla^{2}G^{(2)}({\bf r} )$.
The third term of (\ref{2}) yields
\begin{align}\label{b3}
&-2\frac{\delta}{\alpha \delta J(\textbf{r}^{\prime})}
\left( \frac{\nabla\langle  \psi \rangle}{\langle  \psi \rangle^{2}}\cdot
\langle(\nabla\delta\psi)\delta\psi\rangle \right) |_{J=0}\nonumber\\
&=-\frac{\delta}{\alpha \delta J(\textbf{r}^{\prime})}
\left( \frac{\nabla\langle  \psi \rangle}{\langle  \psi \rangle^{2}}\cdot
\langle\nabla(\delta\psi)^{2}\rangle \right) |_{J=0}\nonumber\\
&=-\frac{1}{\psi_{0}^{2}}\nabla G^{(2)}(0)\cdot\nabla G^{(2)}(\textbf{r}-\textbf{r}^{\prime}).
\end{align}
The fourth term of (\ref{2}) yields
\be \label{b4}
\frac{\delta}{\delta J(\textbf{r}^{\prime})}
\bigg(\frac{(\nabla \langle  \psi \rangle)^{2}}{\langle  \psi \rangle ^{3}}
\langle(\delta\psi)^{2}\rangle\bigg)|_{J=0} =0
\ee
by $\nabla \langle \psi  \rangle=0$.
The sum of Eq. (\ref{2.a+2.b}), (\ref{b3}) and (\ref{b4}) gives
the contribution of the second term of (\ref{main})
\begin{align} \label{2b2}
&-\frac{\delta}{\alpha \delta J(\textbf{r}^{\prime})}\langle\frac{(\nabla\psi)^{2}}{\psi}\rangle|_{J=0}\nonumber\\
=& -\frac{1}{2}\nabla^{2} G^{(2)}(\textbf{r}-\textbf{r}^{\prime})
-k_{J}^{2}\psi_{0} G^{(2)}(\textbf{r}-\textbf{r}^{\prime})
+\frac{\nabla^{2}G^{(2)}(0)}{4\psi_{0}^{2}} G^{(2)}(\textbf{r}-\textbf{r}^{\prime})
+\frac{1}{\psi_{0}^{2}}\nabla G^{(2)}(0)\cdot\nabla G^{(2)}(\textbf{r}-\textbf{r}^{\prime})\nonumber\\
&-\big(\frac{1}{4\psi_{0}}\nabla^{2}+\frac{3}{2}k_{J}^{2}\big)G^{(3)}(\textbf{r},\textbf{r},\textbf{r}^{\prime})
 -\frac{1}{2\alpha }\big( \, \psi_{0}^{2}+3G^{(2)}(0) \, \big)\delta^{(3)}(\textbf{r}-\textbf{r}^{\prime}).
\end{align}

Plugging  Eq. (\ref{nablaG}), (\ref{3b2}), (\ref{4b2}), and (\ref{2b2})
into Eq. (\ref{main}),
we  obtain the equation of 2-point correlation function:
\begin{align} \label{25}
&(\nabla^{2}+2k_{J}^{2}\psi_{0}) G^{(2)}(\textbf{r}-\textbf{r}^{\prime})
+\frac{1}{2\psi_{0}^{2}}\nabla^{2}G^{(2)}(0) G^{(2)}(\textbf{r}-\textbf{r}^{\prime})
-(\frac{1}{2\psi_{0}}\nabla^{2}+k_{J}^{2})G^{(3)}(\textbf{r},\textbf{r},\textbf{r}^{\prime})\nonumber\\
&+\frac{2}{\psi_{0}^{2}}\nabla G^{(2)}(0)\cdot\nabla G^{(2)}(\textbf{r}-\textbf{r}^{\prime})
=-\frac{1}{\alpha }\big( \, \psi_{0}^{2}-G^{(2)}(0) \, \big)\delta^{(3)}(\textbf{r}-\textbf{r}^{\prime}).
\end{align}
This is just Eq. (\ref{2pt3pt}).

Observe that  this equation  is not closed for  $G^{(2)}$,
since it  contains   $G^{(3)}$.
One would go on to get the field equation of  $G^{(3)}$, etc.
This kind of hierarchy is common to the field equation of correlations
in a nonlinear theory when the perturbation method is used.
To cut off the hierarchy and get a closed equation for  $G^{(2)}$,
we adopt the Kirkwood-Groth-Peebles ansatz
in  Eq. (\ref{Ansatz}) (\cite{Kirkwood(1932),GrothPeebles(1977)}).
For simplicity, taking $\textbf{r}^{\prime}$ be the origin $0$,
the ansatz is
\be\label{ansatz}
G^{(3)}(\textbf{r},\textbf{r},\textbf{r}^{\prime})
	=G^{(3)}(\textbf{r},\textbf{r},0)
	=Q \big( 2G^{(2)}(0)G^{(2)}(\textbf{r})+(G^{(2)}(\textbf{r}))^{2} \big).
\ee
It should be noticed  that $G^{(3)}$ is of order $(\delta \psi)^3$
and $G^{(2)}$ is of order $(\delta \psi)^2$,
and the use of ansatz causes an increase of order of  perturbation.
By (\ref{ansatz}),  one has
\begin{align}
-&(\frac{1}{2\psi_{0}}\nabla^{2}+k_{J}^{2})G^{(3)}(\textbf{r},\textbf{r},0)\nonumber\\
=-Q&\bigg(\frac{1}{\psi_{0}}G^{(2)}(0)\nabla^{2}G^{(2)}(\textbf{r})
+\frac{1}{\psi_{0}} (\nabla G^{(2)}(\textbf{r}))^{2}
+\frac{1}{\psi_{0}}G^{(2)}(\textbf{r})\nabla^{2}G^{(2)}(\textbf{r})
+2k_{J}^{2}G^{(2)}(0)G^{(2)}(\textbf{r}) \nonumber\\
&+k_{J}^{2}(G^{(2)}(\textbf{r}))^{2}
+\frac{1}{\psi_{0}}G^{(2)}(\textbf{r})\nabla^{2}G^{(2)}(0)
+\frac{2}{\psi_{0}}\nabla G^{(2)}(0)\cdot \nabla G^{(2)}(\textbf{r})
\bigg).
\label{3in2}
\end{align}
Substituting  (\ref{3in2}) into (\ref{25}), we get
the field equation of $G^{(2)}(\textbf{r})$
\begin{align}
&( \, 1-\frac{Q}{\psi_{0}}G^{(2)}(0) \, )\nabla^{2}G^{(2)}(\textbf{r})
+\bigg(\frac{1}{2\psi_{0}^{2}}(1-2Q\psi_{0})\nabla^{2}G^{(2)}(0)
+2k_{J}^{2}\psi_{0}(1-\frac{Q}{\psi_{0}}G^{(2)}(0)) \bigg) G^{(2)}(\textbf{r})\nonumber\\
&-Qk_{J}^{2}(G^{(2)}(\textbf{r}))^{2}
-\frac{Q}{\psi_{0}}G^{(2)}(\textbf{r})\nabla^{2}G^{(2)}(\textbf{r})
-\frac{Q}{\psi_{0}} (\nabla G^{(2)}(\textbf{r}))^{2}
+\frac{2}{\psi_{0}^{2}}(1-Q\psi_{0})\nabla G^{(2)}(0)\cdot \nabla G^{(2)}(\textbf{r})\nonumber\\
&=-\frac{1}{\alpha } \big( \,  \psi_{0}^{2}-G^{(2)}(0)  \, \big)\delta^{(3)}(\textbf{r}),
\label{G2}
\end{align}
which is closed for $G^{(2)}(\textbf{r})$.
Now we introduce the notations
\begin{align}  \label{k_0}
&\textbf{a}\equiv \frac{2}{\psi_{0}^{2}}(1-Q\psi_{0})\nabla G^{(2)}(0),\\
&b\equiv \frac{Q}{\psi_{0}},\label{bdef}\\
&c\equiv \frac{Qk_{J}^{2}}{k_{0}^{2}},\label{cdef} \\
&k_{0}^{2}\equiv 2k_{J}^{2}\psi_{0}(1-bG^{(2)}(0))
+\frac{1}{2\psi_{0}^{2}}(1-2Q\psi_{0})\nabla^{2}G^{(2)}(0),
\end{align}
then (\ref{G2}) becomes
\begin{align} \label{coeff2}
&\big( \, 1-b G^{(2)}(0)-bG^{(2)}(\textbf{r}) \, \big) \nabla^{2}G^{(2)}(\textbf{r})
	+k_{0}^{2}( \, 1 -cG^{(2)}(\textbf{r}) \, )G^{(2)}(\textbf{r}) \nonumber\\
&+\big( \,  {\bf a}-b\nabla G^{(2)}({\bf r})  \, \big) \cdot\nabla G^{(2)}(\textbf{r})
=-\frac{1}{\alpha }\big( \,  \psi_{0}^{2}-G^{(2)}(0) \,  \big)\delta^{(3)}(\textbf{r}).
\end{align}
Let us do renormalization.
The first term on l.h.s of (\ref{coeff2}) has
$Z_0  \equiv 1-b G^{(2)}(0)$
as part of the coefficient of $\nabla^{2}G^{(2)}(\textbf{r})$,
which can be absorbed in the definition of $G^{(2)} (\bf r)$.
Since $G^{(2)} \propto (\delta\psi)^2$,
this amounts to the renormalization of the density field $\delta \psi$.
Explicitly, multiplying Eq. (\ref{coeff2}) by $Z_0^{-2}$
and making the following substitutions
\begin{align}
& G^{(2)}({\bf r}) \rightarrow G^{(2)}_R({\bf r}) \equiv Z_0^{-1} G^{(2)}({\bf r}),
\hspace{.5cm} k_0^2 \rightarrow k_{0R}^2 \equiv Z_0^{-1} k_0^2, \nonumber \\
& c \rightarrow c_R \equiv Z_0 c,
\hspace{.5cm} {\bf a} \rightarrow {\bf a}_R \equiv Z_0^{-1} {\bf a},\\
&\psi_0^2
\rightarrow \psi_{0R}^2 \equiv Z_0^{-2} (\psi_0^2-G^{(2)}(0)) ,
\end{align}
one  finally obtains
\begin{align} \label{renormeq}
( \, 1-bG^{(2)}(\textbf{r}) \, )\nabla^{2}G^{(2)}(\textbf{r})
+k_{0}^{2}( \, 1- cG^{(2)}(\textbf{r}) \, )G^{(2)}(\textbf{r})
+( \, {\bf a}-b\nabla G^{(2)}({\bf r}) \, ) \cdot\nabla G^{(2)}(\textbf{r})
=-\frac{1}{\alpha }\psi_0^2 \delta^{(3)}(\textbf{r}),
\end{align}
where the subscript  ``$R$" are dropped for simple notations,
and the quantities ${\bf a}$, $c$, $k_0^2$, $G^{(2)}(\bf r)$, and $\psi_0^2$
are understood as the renormalized ones.
The renormalized characteristic wavenumber
is given by $k^2_0=  2 k_{JR}^{2} =8\pi G m_R^2n_0/T$,
where
\be
m_R^2\equiv  m^2\psi_0 + \frac{T}{8\pi G n_0}
Z_0^{-1}\frac{1}{2\psi_{0}^{2}}(1-2Q\psi_{0})\nabla^{2}G^{(2)}(0)
\ee
is the renormalized mass in place of the ``bare"  mass $m$.
Thus, by the renormalization  procedure,
$\nabla^{2}G^{(2)}(0)$ has been absorbed by $m$,
$\nabla    G^{(2)}(0)$ absorbed by $\bf a$,
and $G^{(2)}(0)$ by $\psi_0$ through the multiplicative factor  $Z_0 $.
After the renormalization,
one can set  $\psi_0=1$
in Eq. (\ref{renormeq}).

The parameters   $\bf a$, $b$, $c$  will be regarded as independent parameters.
When ${\bf a}= b= c=0$,
Eq. (\ref{renormeq}) reduces to the Helmholtz equation in Eq. (\ref{Helmoltz})
in the Gaussian approximation.
Thus, the terms involving ${\bf a}$, $b$ and $c$
represent  the nonlinear effects at the order $(\delta\psi)^2$
beyond the Gaussian approximation.
By isotropy of the system,
it is simpler to write the field equation (\ref{renormeq}) in the radial direction.
Denoting
$\xi(x) \equiv G^{(2)}(\textbf{r})$, $x\equiv k_{0}r$,
and
$\xi'\equiv \frac{d}{dx}\xi(x)$,
Eq. (\ref{renormeq}) becomes
\begin{align} \label{equation}
(1-b\xi)\xi''+
( (1-b\xi)\frac{2}{x}+a )\xi' +\xi  -b \xi'\, ^{2} -c\xi^{2}=
-\frac{1}{\alpha }    \frac{\delta(x)k_0}{x^{2}}
\end{align}
where $a\equiv |{\bf a}|/k_0 $.
The dimensionless parameters $a$, $b$, and $ c$ will be regarded
as independent,
as they essentially come from  combinations of
$Q$,  $\nabla G^{(2)}(0)$, and $\nabla^2G^{(2)}(0)$.
The parameter $a$ in Eq. (\ref{equation}) plays a role of the effective viscosity.
The nonlinear terms  $\xi'^2$ and $\xi^2$
can enhance the correlation at small scales.
Compared with Eq. (8) in Ref.(\cite{ZhangMiao(2009)}),
now Eq. (\ref{equation}) has a new term $ -b\xi$
in the coefficients of $ \xi''$  and  $\frac{2}{x}$,
and a new term $-c\xi ^2$.
As we have checked,
the numerical solutions of the two equations differ only slightly
in confronting the observational data.

\section{3-point Correlation  Function in Gaussian Approximation}
\label{App:3PTCF}

Now we derive the field  equation for the 3-point correlation  function $G^{(3)}(\bf r,r',r'')$.
Taking double functional differentiation of Eq. (\ref{masseq})
\be
\frac{\delta^2}{ \delta J({\bf r} ^{\prime}) \delta J({\bf r}'')}
\langle  \nabla^{2}\psi({\bf r} )-\frac{1}{\psi({\bf r} )}(\nabla\psi({\bf r} ))^{2}
+k_{J}^{2}\psi({\bf r} )^{2}+J({\bf r})\psi({\bf r} )^{2}\rangle_J=0,
\ee
and setting $J=0$, one has
\ba\label{eqG3}
&&\nabla ^2_{r}   G^{(3)} ( {\bf r,r',r'' } )
+  2 k_J ^2   \psi_0  G^{(3)} ( {\bf r,r',r'' } )
-  \frac{2}{\psi_0} \nabla  G^{(2)} ( {\bf r,r''}) \cdot \nabla  G^{(2)} ( {\bf r,r'} ) \nn \\
&&+  2k_J ^2   G^{(2)} ( {\bf r,r'} ) G^{(2)} ( {\bf r,r''} )
+  \frac{2}{\alpha}    \psi_0 \delta^{(3)}( {\bf r-r''} ) G^{(2)} ({\bf r,r'})
+     \frac{2}{\alpha}  \psi_0 \delta^{(3)}({\bf r-r'})  G^{(2)} ({\bf r,r''} )  =0 , \nn \\
\ea
where
$\frac{1}{\psi(\bf r)}  \simeq \frac{1}{ \langle  \psi \rangle_J }$
as an approximation  has been used.
This equation is  the field equation of $G^{(3)}$
at  the Gaussian approximation.
To look for its solution, let $G^{(3)}  $
be of the  form (\ref{Ansatz})  of the  anzats,
where  $Q$ is taken as a  constant.
By applying $\nabla  ^2$  on Eq. (\ref{Ansatz})
and plugging into the left hand side of  (\ref{eqG3}) gives
\bee  \label{3ptQ}
& Q &  \bigg( G^{(2)} ({\bf r,r''} )\nabla ^2 G^{(2)} ({\bf r,r'})
       + G^{(2)} ( {\bf r,r' } ) \nabla ^2 G^{(2)} ( {\bf r,r''} )
       + 2 \nabla   G^{(2)} ({\bf r,r'} ) \cdot \nabla   G^{(2)} ({\bf r,r''})  \nonumber \\
& & + \nabla ^2 G^{(2)} ( {\bf r',r} ) G^{(2)} ( {\bf r',r'' } )
    + G^{(2)} ( {\bf r'',r'} )\nabla ^2 G^{(2)} ( {\bf r'',r} )  \bigg)  \nonumber \\
& + & 2 k_J ^2   \psi_0  Q \bigg( \,  G^{(2)} ( {\bf r,r'} ) G^{(2)} ( {\bf r,r'' })
     + G^{(2)} ( {\bf r',r} ) G^{(2)} ({\bf r',r''} )
     + G^{(2)} ( {\bf r'',r'} ) G^{(2)} ( {\bf r'',r} ) \bigg)  \nonumber  \\
& - & \frac{2}{\psi_0}\nabla  G^{(2)} ( {\bf r,r''}) \cdot \nabla  G^{(2)}( {\bf r,r'})
      + 2k_J ^2 G^{(2)} ({\bf r,r'} ) G^{(2)} ( {\bf r,r''} )  \nonumber  \\
& + &  2\delta^{(3)}( {\bf r-r''} ) \frac{1}{\alpha}  \psi_0  G^{(2)} ( {\bf r,r'} )
     + 2\delta^{(3)}( {\bf r-r'} ) \frac{1}{\alpha} \psi_0  G^{(2)} ({\bf r,r''} )  = 0 .
\een
Now we  choose the constant
\[
Q= 1/\psi_0.
\]
Then $Q= 1$
since $\psi_0 =1$ by $\langle \rho\rangle =\rho_0$.
Thus  the two  terms $\nabla G^{(2)}  \cdot \nabla G^{(2)}$ in Eq. (\ref{3ptQ}) cancel,
and one ends up with
\bee \label{re}
& &  G^{(2)} ( {\bf r',r''} ) \nabla ^2 G^{(2)} ( {\bf r',r})
+ G^{(2)} ( {\bf r'',r'} )\nabla ^2 G^{(2)} ( {\bf r'',r})    \nonumber \\
& + &  2k_J ^2 G^{(2)} ({\bf r',r''} )  G^{(2)} ( {\bf r',r} )
+ 2k_J ^2 G^{(2)} ({\bf r'',r'}) G^{(2)} ( {\bf r'',r} )  \nonumber  \\
& + & \frac{1}{\alpha}  G^{(2)} ({\bf r,r''})  \delta^{(3)}({\bf r-r'} )
+ \frac{1}{\alpha}  G^{(2)} ({\bf r,r'})   \delta^{(3)}({\bf r-r''})   \nonumber  \\
& + & G^{(2)} ({\bf r,r''} )\big( \, \nabla ^2 G^{(2)} ({\bf r,r'})
+ 2 k_J ^2 G^{(2)} ({\bf r,r'} ) + \frac{1}{\alpha}   \delta^{(3)}({\bf r-r'}) \, \big) \nonumber \\
&  +   & G^{(2)} ({\bf r,r'}) \big( \, \nabla ^2 G^{(2)} ({\bf r,r''} )
+ 2k_J ^2  G^{(2)} ({\bf r,r''}) + \frac{1}{\alpha}   \delta^{(3)}({\bf r-r''}) \, \big) = 0 .
\een
By the field equation (\ref{Helmoltz})
of   $G^{(2)} $ at Gaussian approximation,
the last two terms vanish, and     (\ref{re})   reduces to
\be \label{eqf}
\frac{1}{\alpha}  \delta^{(3)}({\bf r-r''}) \big( \, G^{(2)} ({\bf r,r'})-G^{(2)} ({\bf r',r''}) \, \big)
+ \frac{1}{\alpha} \delta^{(3)}( {\bf r-r'} ) \big( \, G^{(2)} ({\bf r,r''}) -G^{(2)} ({\bf r',r''}) \, \big)  = 0.
\ee
By the property of  $\delta$-function
and by  $G^{(2)} ({\bf r,r'})= G^{(2)} ({\bf r',r}) =G^{(2)}(|{ \bf r-r'}|)$,
the equation (\ref{eqf})  is satisfied automatically.
Thus,
at the level of the Gaussian approximation  of our theory,
the Kirkwood-Groth-Peebles ansatz  (\ref{Ansatz})  with $Q= 1$
holds exactly as a relation  between
$G^{(2)}$  of  Eq. (\ref{2ptsol})
and  $G^{(3)}$ of  Eq. (\ref{3ptsol}).

\end{document}